\documentclass[bibyear]{aa}

\usepackage{graphicx}
\usepackage{natbib}  
\usepackage{float}
\usepackage{color}
\usepackage{txfonts}
\usepackage{gensymb}
\usepackage{amsmath}
\usepackage{xcolor}
\usepackage{xspace}
\usepackage{upgreek}
\usepackage[normalem]{ulem}
\usepackage[colorlinks=true,citecolor=blue,linkcolor=blue,urlcolor=blue,breaklinks=true]{hyperref}
\newcommand{\src}{FRB\,20201124A}
\newcommand{\srcRi}{FRB\,20121102A}
\newcommand{\srcRiii}{FRB\,20180916B}
\newcommand{\dmu}{pc\,cm$^{-3}$\xspace}
\newcommand{\bandiii}{band~3\xspace}
\newcommand{\bandiv}{band~4\xspace}

\newcommand{\am}{\alpha_\mathrm{m}}
\newcommand{\at}{\alpha_\mathrm{t}}
\newcommand{\Ns}{N}
\newcommand{\FRBperiodic}{FRB\,20191221A}
\newcommand{\scbw}{$\Delta\nu_{\text{sc}}$}
\newcommand{\Spring}{S21\xspace}
\newcommand{\Fall}{F21\xspace}
\newcommand{\Winter}{W22\xspace}

\title{An activity transition in FRB 20201124A:\\
  Methodological rigor, 
  detection of frequency-dependent cessation,
  and a geometric  magnetar model.
}

\authorrunning{Bilous et al.}
\titlerunning{Frequency-dependent activity in FRB 20201124A}

\author{%
  A.~V.~Bilous\inst{1,2}\thanks{ORCID: \href{https://orcid.org/0000-0002-7177-6987}{0000-0002-7177-6987}} \and
  J.~van~Leeuwen\inst{1}\thanks{ORCID: \href{https://orcid.org/0000-0001-8503-6958}{0000-0001-8503-6958}} \and
  Y.~Maan\inst{3,1}\thanks{ORCID: \href{https://orcid.org/0000-0002-0862-6062}{0000-0002-0862-6062}} \and
  I.~Pastor-Marazuela\inst{4,1,5}\thanks{ORCID: \href{https://orcid.org/0000-0002-4357-8027}{0000-0002-4357-8027}} \and
  L.~C.~Oostrum\inst{1,4,6}\thanks{ORCID: \href{https://orcid.org/0000-0001-8724-8372}{0000-0001-8724-8372}} \and
  K.~M.~Rajwade\inst{7}\thanks{ORCID: \href{https://orcid.org/0000-0002-8043-6909}{0000-0002-8043-6909}} \and
  Y.~Y.~Wang\inst{4}\thanks{ORCID: \href{https://orcid.org/0000-0002-3822-0389}{0000-0002-3822-0389}}
}

\institute{ASTRON, the Netherlands Institute for Radio Astronomy, Oude Hoogeveensedijk 4,7991 PD Dwingeloo, The Netherlands\label{astron}
  \and
Independent researcher, Zwolle, The Netherlands\label{indep}, \email{hanna.bilous@gmail.com}
  \and
National Centre for Radio Astrophysics, Tata Institute of Fundamental Research, Pune 411007, Maharashtra, India\label{ncra}
  \and
Anton Pannekoek Institute, University of Amsterdam, Postbus 94249, 1090 GE Amsterdam, The Netherlands\label{uva}
  \and
Jodrell Bank Centre for Astrophysics, Department of Physics and Astronomy, The University of Manchester, Manchester, M13
9PL, UK\label{jbca}
  \and
Netherlands eScience Center, Science Park 402, 1098 XH Amsterdam, The Netherlands\label{escience}
  \and
Astrophysics, University of Oxford, Denys Wilkinson Building, Keble Road, Oxford, OX1 3RH, United Kingdom\label{oxford}
}

\abstract{We report detections of fast radio bursts (FRBs) from the 
 repeating source \src\  with Apertif/WSRT and GMRT, and measurements 
 of basic burst properties, especially the  dispersion measure (DM) 
 and fluence. Based on comparisons of these properties with 
 previously published larger samples, we argue that the excess 
 DM reported earlier for pulses with integrated signal-to-noise ratios 
 $\lesssim 1000$  is due to incompletely
accounting for what is known as the sad trombone effect, even when using 
structure-maximizing DM algorithms. Our investigations of fluence 
distributions next lead us to advise against formal  power-law fitting; 
we especially caution against the use of the least-squares method, and 
we demonstrate the large biases involved.
A maximum likelihood estimator (MLE) provides a much more accurate estimate 
of the power law, and we provide accessible code for direct inclusion in future research.
Our GMRT observations were fortuitously 
scheduled around the end of the {Spring 2021} activity window
as recorded by FAST. 
We detected several bursts (one of them very 
strong) at  400/600\,MHz, a few hours after sensitive FAST 
non-detections already showed the 1.3\,GHz FRB emission to have ceased.
After FRB\,20180916B, this is a second example of a frequency-dependent 
activity window identified in a repeating FRB source. Since numerous 
efforts have  so far failed to determine a spin period for FRB\,20201124A,  
we conjecture that it is an ultra-long-period  magnetar, with a 
period on the scale of months, and with a very wide, highly irregular 
duty cycle. Assuming the emission comes from closed field lines, 
we used radius-to-frequency mapping and polarization information
from other studies to constrain the magnetospheric geometry and 
location of the emission region. Our initial findings are consistent
with a possible connection between FRBs and crustal motion 
events.
 }
   

\begin{document} 

\maketitle

\section{Introduction}
Fast  radio bursts (FRBs) are microsecond- to millisecond-long bursts of radio emission 
of extragalactic origin \citep[see][for review]{Petroff2019,Petroff2022}. 
Over the almost two decades since the discovery of the first FRB 
\citep{Lorimer2007}, multiple  FRB emission theories have been proposed 
\citep[for the live catalog see][]{Platts2019}. To date no consensus 
about their exact origin has emerged, although neutron star progenitors currently 
appear to be favored.

It is even possible that several classes of progenitors emit FRBs. At the 
moment, the most obvious empirical demarcation between these types is the 
dichotomy between one-off sources and FRB repeaters. Only the former are 
potentially cataclysmic. These two classes show statistical differences 
in the spectro-temporal  properties of the bursts \citep{Pleunis2021a}.  
In comparison to one-off sources, repeaters offer much more information 
about their circum-burst environment  and the burst emission mechanism: they can 
be localized with much better precision and the pulses provide dynamical 
estimates of the electron content and magnetic field in the vicinity of 
the emitting plasma. Any emission theory must explain the entire distribution 
of burst fluences, and the various spectro-morphological properties of pulses 
that originate in the same local environment.  

While some FRBs are only seen to repeat a handful of times, \src\ is a 
veritable  FRB factory: it is capable of emitting prolifically, and 
datasets covering it contain hundreds of pulses. It is thought to be 
located in a dynamically evolving magnetized environment, as suggested 
by irregular rotation measure (RM) variations on short timescales 
and the presence of Faraday conversion \citep{Xu2021}. \src\ is 
notorious for its high but exceedingly variable pulse emission rate. 
The source was first detected at the end of 2020 by the Canadian Hydrogen
Intensity Mapping Experiment (CHIME), but only after about 40\,hours of 
earlier observations of the same field contained no detections  \citep{Lanman2022}.
By 2021 March-May \src\ had entered a high-activity phase, hereafter called  the Spring 2021 
(\Spring) epoch,\footnote{Given the source declination of 
$+26\degree$ we label the activity spans  by the northern hemisphere seasons.}
reaching rates almost 50 bursts per hour, as observed by the Five-hundred-meter 
Aperture Spherical radio Telescope (FAST) at 20\,cm wavelength \citep{Xu2021}.
The \Spring\ FAST burst sample was complemented by observations performed 
with CHIME, the Effelsberg 100 m and the Parkes 64 m  dishes, the upgraded 
Giant Metrewave Radio Telescope (uGRMT), and  the Australian Square 
Kilometre Array Pathfinder (ASKAP) \citep{Lanman2022, Hilmarsson2021, Marthi2022, Kumar2022}. 

Bursts from \src\ exhibit a high degree of circular and linear polarizations, 
with predominantly flat position angle (PA) curves. These polarization 
properties hint at a magnetospheric origin \citep{Jiang2022}. Such an origin 
agrees with a rotating neutron-star progenitor hypothesis. However, despite 
extensive searches, no periodicity has been found in the high number of bursts, 
over a broad range of trial periods spanning from milliseconds to days 
\citep{Niu2022,Du2023}.

The FAST observations indicate that the \Spring\ activity epoch  ended abruptly 
between 2021 May 26 and 29 \citep[][observations at 1250\,MHz]{Xu2021}. 
On May 27 CHIME/FRB recorded one more burst from this FRB at the 
lower frequencies of 400--800\,MHz \citep{Lanman2022}, 
followed by a bright burst at 1350\,MHz detected by the Stockert telescope 
on May 28, during a three-day gap in FAST coverage \citep{Kirsten2024} .
The gaps in the observing schedules of these three telescopes did not 
allow further constraints on any possible frequency-dependent boundaries of 
the activity phase, an effect seen in one other repeating FRB, \srcRiii\ 
\citep{Pastor-Marazuela2021}. The end of the {activity window} was, however, 
also covered by GMRT observations at 300--750\,MHz, {which are 
presented in this work}. Strikingly, GMRT 
detected several low-frequency pulses on June\,1, several hours after 
the FAST non-detections showed the higher-frequency bursts had already 
turned off.

Despite continued monitoring \citep{Mao2022, Trudu2022}, these June\,1 
pulses were the only signs of activity detected from \src\, until 
the pulses reappeared at some time before 2021 September 21  \citep[][also
  CHIME\footnote{\url{https://www.chime-frb.ca/repeaters/FRB20201124A}}]{Main2021}.
An extensive FAST monitoring campaign then observed an exponentially increasing 
FRB rate, reaching activity levels an order of magnitude higher than those 
\citeauthor{Xu2021}  reported earlier. The  activity in this Fall 2021 
(\Fall) epoch abruptly quenched again between September 28 and 29 
\citep{Zhou2022}. No detections were   published  until late 2022 January 
\citep[the ``Winter 2022'', W22 epoch;][]{Ould-Boukattine2022}.

The repeater field was  regularly scheduled in the 
Apertif-LOFAR\footnote{The APERture Tile In Focus and LOw-Frequency ARray, 
respectively.} Exploration of the Radio Transient sky (ALERT) survey, 
that started in 2019 at the Westerbork Synthesis Radio Telescope 
\citep[WSRT;][]{Maan2017,vanLeeuwen2022,Pastor-Marazuela2024}.
During the last observing run of this survey, in the first week of 2022 
February, the three observing sessions toward \src\  yielded ten FRB 
detections.

This work describes the properties of those bursts detected within the 
ALERT survey in \Winter, and of GMRT bursts from the \Spring and \Winter
{activity window}s. We include the analysis of dispersion measures (DMs), 
fluence distributions, \mbox{(quasi-)}periodicities, and the scintillation of 
these bursts. 

The evidence we find for a frequency-dependent {activity window}, together with the 
energetics and morphological properties of the recorded pulses -- both 
in our sample and in the large FAST sample -- provide  a unique  
test of the hypothesis that FRBs originate from low-twist ultra-long-period 
(ULP) magnetars  \citep{Wadiasingh2019,Wadiasingh2020a,Beniamini2020,Caleb2022}. 
That is interesting because a variety of scenarios have been proposed to explain how neutron stars 
might form FRBs, but testing and distinguishing these observationally is challenging.

In the ULP hypothesis, FRBs are generated via a pulsar-like emission mechanism 
in the magnetospheres of very slowly rotating magnetars. In these sources,
the non-potential magnetosphere (i.e., one in which currents flow)
is characterized by an unusually weak large-scale magnetic field twist, 
much weaker than that of commonly observed galactic magnetars. This results in 
a low plasma density on the closed magnetic field lines. In such 
charge-starved conditions, deformations in the crust that dislocate the 
footpoints of the magnetic field lines generate strong transient electric 
fields. This, in turn, leads to avalanche pair production and, ultimately, 
the emission of FRBs via a pulsar-like coherent mechanism.

The low-twist magnetar hypothesis of FRB generation makes specific predictions 
for the  times of arrival (TOAs) of individual bursts, and for the 
quasi-periodicity of  sub-bursts. In this work, we  compare these predictions to 
existing observational evidence. Finally, we  put constraints on the 
location and shape of active regions on the surface of the star by using two 
classical phenomenological models of radio pulsar emission: radius-to-frequency 
mapping and rotating vector models.

%%%%%%%%%%%%%%%%%%%%%     GMRT observations
\section{GMRT observations and analysis}
\label{sect:GMRT}

\begin{table}[b]
\begin{center} 
\caption{FRBs detected with GMRT.}
\begin{tabular}{p{0.4cm}p{0.4cm}cccp{0.4cm}p{0.4cm}p{0.4cm}} 
\hline\\ %[0.01cm]
\parbox{0.5cm}{\centering  Burst\\\#}&
\parbox{0.5cm}{\centering  $\nu_\mathrm{c}$\\(MHz)} &
\parbox{1.5cm}{\centering Topo TOA\\(MJD)} &
\parbox{0.7cm}{\centering  DM\\(pc\\cm$^{-3}$)}&
\parbox{0.7cm}{\centering Equiv. width (ms)} &
\parbox{0.7cm}{\centering Peak Flux (Jy)} &
\parbox{0.7cm}{\centering $F$ (Jy\\ms)} 
\\ [0.3cm]
\hline\\
%# DM DMer w_boxcar  w_acf range speak fluence
G01 &400 &  59366.29432631  & 412.7 & 42 & \phantom{0}3.8 & 160 \\    
G02 &400 &  59366.31507238  & 412.5 & 19 & \phantom{0}2.1 & \phantom{0}40 \\     
G03 &650 &  59366.30413889  & 414.5 & 15 & \phantom{0}3.2 & \phantom{0}48 \\    
G04 &650 &  59366.31848719  & 416.0 & 18 & \phantom{0}4.1 & \phantom{0}72 \\       
G05 &650 &  59366.32968037  & 415.5 & 16 & 43.4 & 679 \\     
G06 &650 &  59366.34033776  & 415.0 & 33 & \phantom{0}4.4 & 144 \\
G07 &650 & 59616.77764372 & 419.8 & 23 & \phantom{0}2.2 & \phantom{0}51 \\
G08 &650 & 59617.55751631 & 417.2 & 11 & \phantom{0}2.6 & \phantom{0}28 \\
G09 &650 & 59617.56061649 & 412.2 & 18 & \phantom{0}1.2 & \phantom{0}22 \\
G10 &650 & 59617.56817339 & 416.4 & 13 & \phantom{0}2.1 & \phantom{0}27 \\
G11 &650 & 59617.57193117 & 421.0 & 29 & \phantom{0}2.1 & \phantom{0}59 \\  
G12 &650 & 59617.59070828 & 416.0 & 20 & \phantom{0}1.8 & \phantom{0}34 \\  
G13 &650 & 59617.55159768 & 418.2 & 18 & \phantom{0}1.4 & \phantom{0}26 \\
G14 &650 & 59617.55230447 & 421.0 & 20 & \phantom{0}1.4 & \phantom{0}28 \\  
G15 &650 & 59617.63759893 & 413.4 & \phantom{0}7 & \phantom{0}1.5 & \phantom{0}11 \\
G16 &650 & 59617.61070096 & 413.6 & 12 & \phantom{0}2.3 & \phantom{0}26 \\
\\[0.1cm]
\hline 
\end{tabular} 
\tablefoot{The columns denote: burst number, central frequency 
$\nu_\mathrm{c}$, TOA at telescope site at center frequency, detection DM, 
equivalent width, peak flux density (dedispersed to $\mathrm{DM}=414.73$\,\dmu), 
and fluence $F$. \label{table:FRBs_GMRT}}
\end{center}
\end{table}

\begin{figure*}
\centering
 \includegraphics[width=0.7\textwidth]{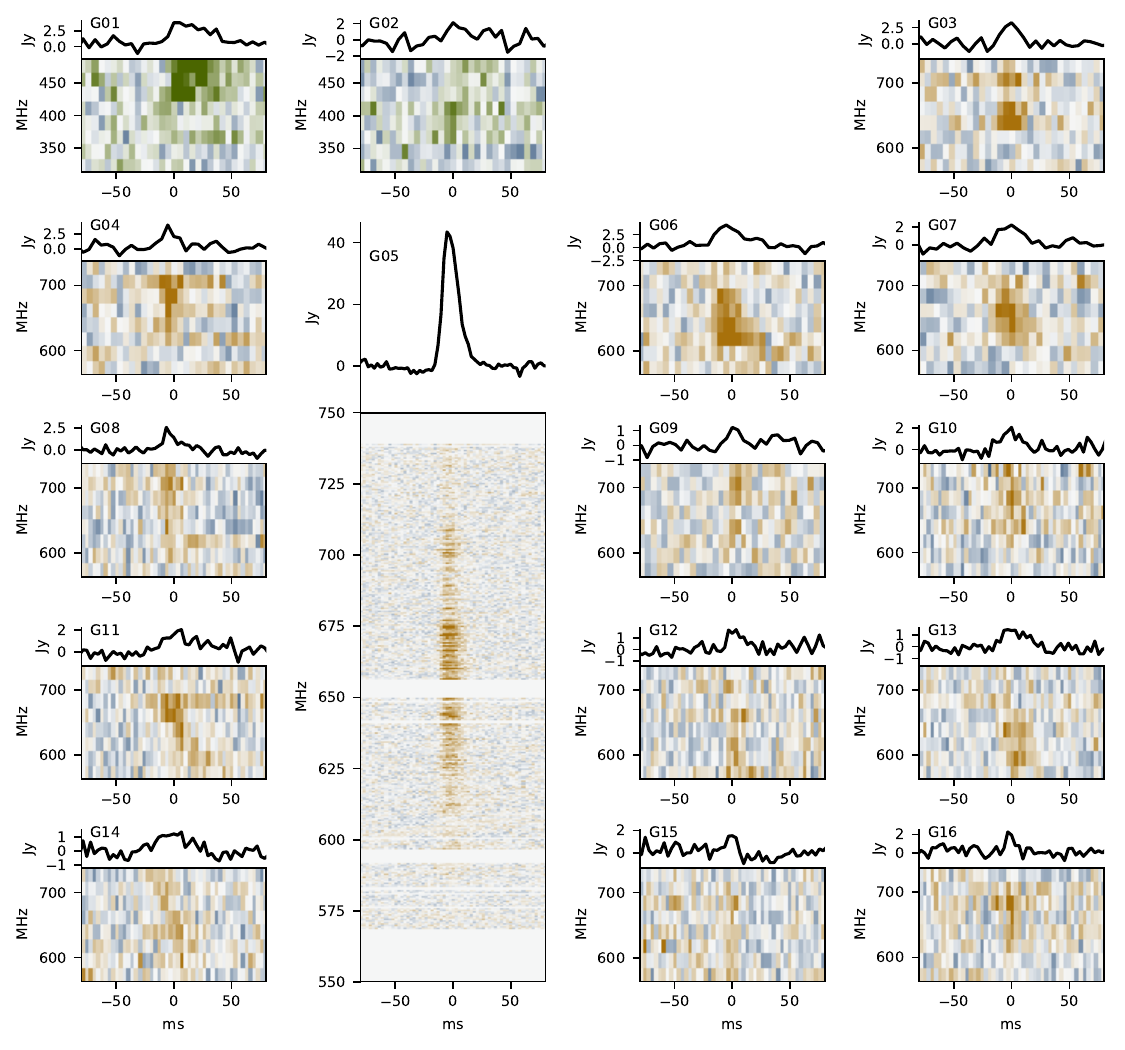}
 \caption{Spectra and band-integrated sample intensity profiles for the 16 
 bursts from GMRT observations. Bursts G01--G06 were detected on  2021 June 01, 
 the rest on 2022 February 06--07. All bursts are dedispersed to $\mathrm{DM}=414.73$\,\dmu\,
 determined from burst G05. For plotting, the spectra were normalized by the mean 
 and standard deviation in the off-burst region in each sub-band and the colors 
 were saturated at $\pm8\sigma$ for G05 and at $\pm4\sigma$ for the other
 bursts. All bursts from 2021 except for G05 are plotted with a time resolution 
 of 5.24\,ms and a frequency resolution of 22--25\,MHz. Burst G05 is shown on a 
 finer 2.62 ms/391 kHz spectro-temporal grid. For the 2022 bursts, the time
 resolution is 3.3-5.9\,ms, and the frequency resolution is 24\,MHz. }
\label{fig:FRBs_GMRT}
\end{figure*}

%%%%%%%%%%%%
\subsection{Observations}

We observed \src\ with the uGMRT \citep{Gupta2017} on 2021 June 01, 
July 03 \& 04, and 2022 February 05,  
06 \& 07, under Director's Discretionary Time. Except for one session, \src\ 
was observed simultaneously in the \bandiii (300--500\,MHz) and \bandiv 
(550--750\,MHz) of GMRT, by combining the antennae in two different sub-arrays. 
These dual-band observations thus provided a frequency coverage of 300--750\,MHz. 
The observing setup utilized the phased-array beam (PAB) sensitivity of typically 
11 antennae in each of the two sub-arrays. At \bandiv, data were recorded at 
4096 channels across 200\,MHz bandwidth centered at 650\,MHz, with a sampling 
time of 0.328\,ms. At \bandiii, the data were coherently dedispersed in real-time 
at a DM of 410.9\,{\dmu}, and then recorded to disk with 
1024 sub-bands across 200\,MHz bandwidth centered at 400\,MHz, using a sampling 
time of 0.164\,ms. The last session on 2022 February 07 was conducted only in 
\bandiv utilizing a higher sensitivity obtained by combining 16 antennae in a 
single sub-array, and the PAB data were coherently dedispersed at a DM of 
410.9\,\dmu in real-time and then recorded to the disk with 1024 sub-bands and 
a sampling time of 40.96\,$\upmu$s. During all the observations conducted in 2022, 
data from the incoherent-array beam (IAB)  formed using the same antennae in the 
individual sub-arrays were also recorded simultaneously with the PAB. The 
availability of the IAB data facilitate a better excision of the radio 
frequency interference (RFI). 

%%%%%%%%%%%%
\subsection{Preprocessing and search for radio bursts}

The individual band data from all the sessions were processed through the following 
series of data reduction steps. For the \bandiv observations, which did not employ the
real-time coherent dedispersion, we used SIGPROC's \texttt{dedisperse} to create 
1024 sub-bands, by dedispersing sets of 4 channels to a single sub-band, using a 
DM of 410.9\,\dmu. As mentioned above, all the \bandiii observations were already 
recorded with 1024 coherently dedispersed sub-bands (the real-time coherent 
dedispersion removes only the intra-sub-band smearing). The individual band data 
from each of the sessions were then subjected to down-sampling from 16\,bits to 
8\,bits per sample using \texttt{digifil}, and RFI excision using 
\texttt{RFIClean}\footnote{\url{https://github.com/ymaan4/RFIClean}} 
\citep{Maan2021} and \texttt{rfifind} from the pulsar search and analysis toolkit 
\texttt{PRESTO} \citep{Ransom2002}.

For the observations conducted in 2022, we formed the post-correlation beam 
\citep{Roy2018} using the PABs and IABs (i.e., we subtracted the IAB from 
the PAB after appropriate scaling\footnote{\url{https://github.com/ymaan4/pcBeam-GMRT}}). 
This post-correlation beam contains less red noise, mitigates some RFI and thus 
reduces false candidates while searching for bright pulses. This beam was processed through
the same steps described above.

The sub-banded data were then incoherently dedispersed in steps of 0.05 and 
0.2\,\dmu, covering the DM range 400--425\,\dmu\ in 500 and 125 trial DMs, for 
\bandiii and \bandiv respectively, using \texttt{prepdata} from \texttt{PRESTO}. 
The above configurations limit the dispersive smearing to a maximum of 1\,ms
throughout the explored DM range for both bands. Each dedispersed time-series 
was searched for the presence of bright pulses above a signal-to-noise ratio
(S/N) threshold of 8 and a maximum pulse-width of 50\,ms using 
\texttt{single\_pulse\_search.py}. All the single pulse candidates were next
grouped by looking for the brightest candidates within roughly 100 or 200\,ms 
wide time windows and across all the trial DMs. Waterfall plots for all the 
grouped candidates were scrutinized by eye to identify the genuine bursts.

Each dedispersed time-series was also searched for periodic signals using 
\texttt{accelsearch} from \texttt{PRESTO}, with \texttt{zmax} set to 256. For 
each observation, all the periodic signal candidates were sifted for harmonics 
and duplicates at different DM trials, and the final candidates were folded 
using \texttt{prepfold} and the diagnostic plots were examined by human experts.

Overall, 16 bursts were detected, six during the 2021 June\,01 session, one 
on 2022 February\,6 and the rest on February\,07. The bursts are labeled 
G01--G16 following their TOA order. Figure~\ref{fig:FRBs_GMRT} 
shows spectra and band-integrated profiles for the obtained burst sample.

{Table~\ref{table:FRBs_GMRT} lists the burst properties. Burst fluences
were determined by integrating the emission over the 100 ms window centered at the burst peak.
The equivalent width was calculated by dividing the fluence by the peak flux density.}

All bursts except G05 are faint and little more can be inferred from their 
spectro-temporal shapes than  a hint of downward-drifting trombone features, 
when dedispersed to the DM obtained from G05 (see Section~\ref{sec:DM}). Overall, 
the peak flux densities, fluences and equivalent widths are comparable to the 
\src\ bursts recorded earlier with GMRT, in 2021 April \citep{Marthi2022}.

%%%%%%%%%%%%%%%%%%%%%%%% ALERT observations
\section{ALERT observations and analysis}
\label{sect:WSRT}

Apertif is a phased-array front-end system installed on 12 of the 14 WSRT
dishes\footnote{Westerbork dish RT1 also monitors  FRBs, in stand-alone mode 
(e.g., \citealt{Kirsten2024}).  We hereafter call that mode Wb-RT1. All other 
references to WSRT imply Apertif.} \citep{Adams2019,vanCappellen2022}.
Apertif observed \src\  as part of the scheduled visits of repeater fields in 
the ALERT survey \citep{Oostrum2020b,vanLeeuwen2022}, on 2021 July 03 \& 04,
and on 2022 February 01, 05, \& 06. The last two observations were coordinated 
to overlap with GMRT (see Sect.~\ref{sect:simult}). All sessions lasted for 
three hours except for a 2.4-hr-long session on February 05.

Apertif consists of phase array feeds on a multi-element interferometer,
that form a hierarchical system of beams. During the \src\, observations, 
the central compound beam CB00 was pointed 
at the J2000 sky coordinates $\mathrm{RA} = 05^\mathrm{h}08^\mathrm{m}03.5^\mathrm{s}$, 
\mbox{$\mathrm{Dec}=+26\degree03\arcmin38.4\arcsec$} ($37.8\arcsec$ for July 2021 sessions). 
These coordinates are close to the source position at 
$\mathrm{RA} = 05^\mathrm{h}08^\mathrm{m}03.5073^\mathrm{s}\pm 4.7\,\mathrm{mas}$,
\mbox{$\mathrm{Dec}=+26\degree03\arcmin38.5032\arcsec\pm3.9\,\mathrm{mas}$} 
as reported by \citet{Nimmo2022}. The offset lies well within the Apertif 
localization limits \citep{Oostrum2020a}.

Total intensity samples were recorded at a  time resolution of 81.92\,$\mu$s, 
and with 1536 channels of 195.312\,kHz for a sample bandwidth of 300\,MHz 
centered at 1369.6\,MHz. As part of the standard real-time FRB search, the 
data from all 40 compound beams was independently searched for FRBs,
using the \mbox{AMBER}\footnote{\url{https://github.com/TRASAL/AMBER}} search code
\citep[the Apertif Monitor for Bursts Encountered in Real-Time;][]{Sclocco2016}
in the DARC\footnote{\url{https://github.com/loostrum/darc}} pipeline
\citep[the Data Analysis of Real-time Candidates;][]{Oostrum2021} on the 
Apertif Radio Transient System \citep[ARTS;][]{vanLeeuwen2014}. Real-time 
candidate selection was carried out by a neural network, as described in 
\citet{Connor2018}.

The DARC pipeline did not find any candidates in the July sessions. This is 
in line with the reports by \citet{Main2021} on targeted observations with
both uGMRT and the Effelsberg Telescope over 2021 June--Aug. \citet{Mao2022} 
also did not find any bursts down to 4\,Jy\,ms during a 104-hr observing run with
Nanshan 26 m Radio Telescope in 2021 June-July. 

During the 2022 February ALERT run, two bright FRBs were detected on the
1st, at $\mathrm{DM}=410$\,\dmu. Initially, the bursts were found in CB17, 
which partially overlaps CB00 \citep[Fig.~3 in][]{vanLeeuwen2022}. Fluences 
and dispersion measures for these bursts were previously reported in 
\citet{Atri2022}. Subsequent reanalysis showed the pulses were not detected
in CB00 because of the residual RFI. The next session, on the February~05, 
yielded one more burst. No bursts were detected on the February~06, despite 
comparable system parameters and RFI. 

%%%%%%%%%%%%%%%%%
\subsection{Deep search}

We  subsequently performed an offline search over a finer grid of trial DMs and 
matched-filter boxcar widths than is possible in the real-time search. First, the 
filterbank files from CB00 were thoroughly cleaned of RFI using the 
\texttt{iqrm}\footnote{\url{https://gitlab.com/kmrajwade/iqrm_apollo}} 
software implementation of the Inter-Quartile Range Mitigation outlier detection 
algorithm \citep{Morello2022}. 
{After cleaning with \texttt{iqrm}, we applied
  \texttt{rficlean},\footnote{\url{https://github.com/ymaan4/RFIClean}, version 2.7.}  which operates in the Fourier
  domain \citep{Maan2021}. We employed the default threshold values, but omitted the excision of spiky RFI in the band-averaged time series.}

Pulse candidates  were selected using 
\texttt{TransientX}\footnote{\url{https://github.com/ypmen/TransientX}}
\citep{Men2024}. 
We searched for FRBs over  DMs ranging from  360 to 460\,\dmu\ with a 
step of 0.25\,\dmu. 
The search was performed on time series at the original time resolution. Boxcar filter widths ranged from 0.18\,ms (slightly 
over two samples) to 200\,ms. Each session yielded about 500 potential 
FRBs with integrated $\mathrm{S/N}\geq7$, which were inspected visually. 
Most of the candidates were residual RFI, displaying sharp, narrow-band 
positive and negative jumps in the frequency-resolved signal. Only 24 
pulses possessed the FRB-like properties we defined, being relatively
broadband and displaying smooth variation of signal strength with frequency. 
Three brighter candidates matched earlier DARC detections.

\begin{table}
\begin{center} 
\caption{Detected FRBs and FRB candidates.}
\begin{tabular}{ccccc} 
\hline\\ %[0.01cm]
\parbox{1.2cm}{\centering  Burst \#}&
\parbox{1.2cm}{\centering  TOA (MJD)}&
\parbox{1.2cm}{\centering DM (\dmu)} &
\parbox{1.2cm}{\centering S/N} &
\parbox{1.2cm}{\centering Width (ms)} 
\\ [0.3cm]
\hline\\
 & \textcolor{gray}{ 59398.32763124} & \textcolor{gray}{432.50} & \textcolor{gray}{\phantom{0}7.1} & \textcolor{gray}{\phantom{0}4.1} \\ 
 & \textcolor{gray}{ 59398.35390255} & \textcolor{gray}{423.00} & \textcolor{gray}{\phantom{0}7.3} & \textcolor{gray}{\phantom{0}3.4} \\ 
 & \textcolor{gray}{ 59399.30523345} & \textcolor{gray}{407.50} & \textcolor{gray}{\phantom{0}7.5} & \textcolor{gray}{\phantom{0}0.7} \\ 
 & \textcolor{gray}{ 59399.33769469} & \textcolor{gray}{377.00} & \textcolor{gray}{\phantom{0}7.3} & \textcolor{gray}{\phantom{0}1.1} \\ 
 & \textcolor{gray}{ 59611.83116861} & \textcolor{gray}{424.00} & \textcolor{gray}{\phantom{0}7.1} & \textcolor{gray}{21.3} \\ 
W01\phantom{0} &  59611.83130447 & 418.50 & 93.3 & \phantom{0}3.4 \\ 
 & \textcolor{gray}{ 59611.83210172} & \textcolor{gray}{369.50} & \textcolor{gray}{\phantom{0}7.7} & \textcolor{gray}{\phantom{0}0.6} \\ 
W02\phantom{0} &  59611.84358984 & 434.00 & 10.5 & 11.4 \\ 
 & \textcolor{gray}{ 59611.84811625} & \textcolor{gray}{416.75} & \textcolor{gray}{\phantom{0}7.1} & \textcolor{gray}{11.4} \\ 
W03\phantom{0} &  59611.86041082 & 422.75 & 12.2 & 32.4 \\ 
W04\phantom{0} &  59611.87300625 & 425.75 & 11.6 & \phantom{0}9.3 \\ 
W05\phantom{0} &  59611.87448743 & 433.50 & 15.8 & 17.3 \\ 
W06\phantom{0} &  59611.89157873 & 428.25 & \phantom{0}9.1 & \phantom{0}9.3 \\ 
 & \textcolor{gray}{ 59611.90475537} & \textcolor{gray}{433.00} & \textcolor{gray}{\phantom{0}7.5} & \textcolor{gray}{21.3} \\ 
 & \textcolor{gray}{ 59611.90700670} & \textcolor{gray}{418.00} & \textcolor{gray}{\phantom{0}7.7} & \textcolor{gray}{14.0} \\ 
W07a &  59611.90761850 & 412.75 & \phantom{0}9.6 & \phantom{0}6.1 \\ 
W07b &  59611.90761951 & 425.50 & 57.3 & 21.3 \\ 
& \textcolor{gray}{ 59611.91905550} & \textcolor{gray}{390.50} & \textcolor{gray}{\phantom{0}7.3} & \textcolor{gray}{\phantom{0}7.5} \\ 
W08\phantom{0} &  59611.92021812 & 422.00 & \phantom{0}9.3 & \phantom{0}7.5 \\ 
W09\phantom{0} &  59611.92725919 & 425.25 & 11.3 & 11.4 \\ 
 & \textcolor{gray}{ 59611.92842737} & \textcolor{gray}{414.25} & \textcolor{gray}{\phantom{0}7.4} & \textcolor{gray}{\phantom{0}3.4} \\ 
W10\phantom{0} &  59615.74587387 & 415.25 & 51.7 & \phantom{0}6.1 \\ 
 & \textcolor{gray}{ 59616.73514411} & \textcolor{gray}{432.00} & \textcolor{gray}{\phantom{0}7.6} & \textcolor{gray}{17.3} \\ 
 & \textcolor{gray}{ 59616.78902312} & \textcolor{gray}{424.75} & \textcolor{gray}{\phantom{0}7.4} & \textcolor{gray}{\phantom{0}9.3} \\ 

\\[0.1cm]
\hline 
\end{tabular}
\tablefoot{The columns are: burst number, 
TOA at the center of observing band at the telescope site, DM, integrated 
S/N, and equivalent width from \texttt{transientX}. \label{table:transientX}}
\end{center}
\end{table}

\begin{figure*}
\centering
 \includegraphics[width=0.7\textwidth]{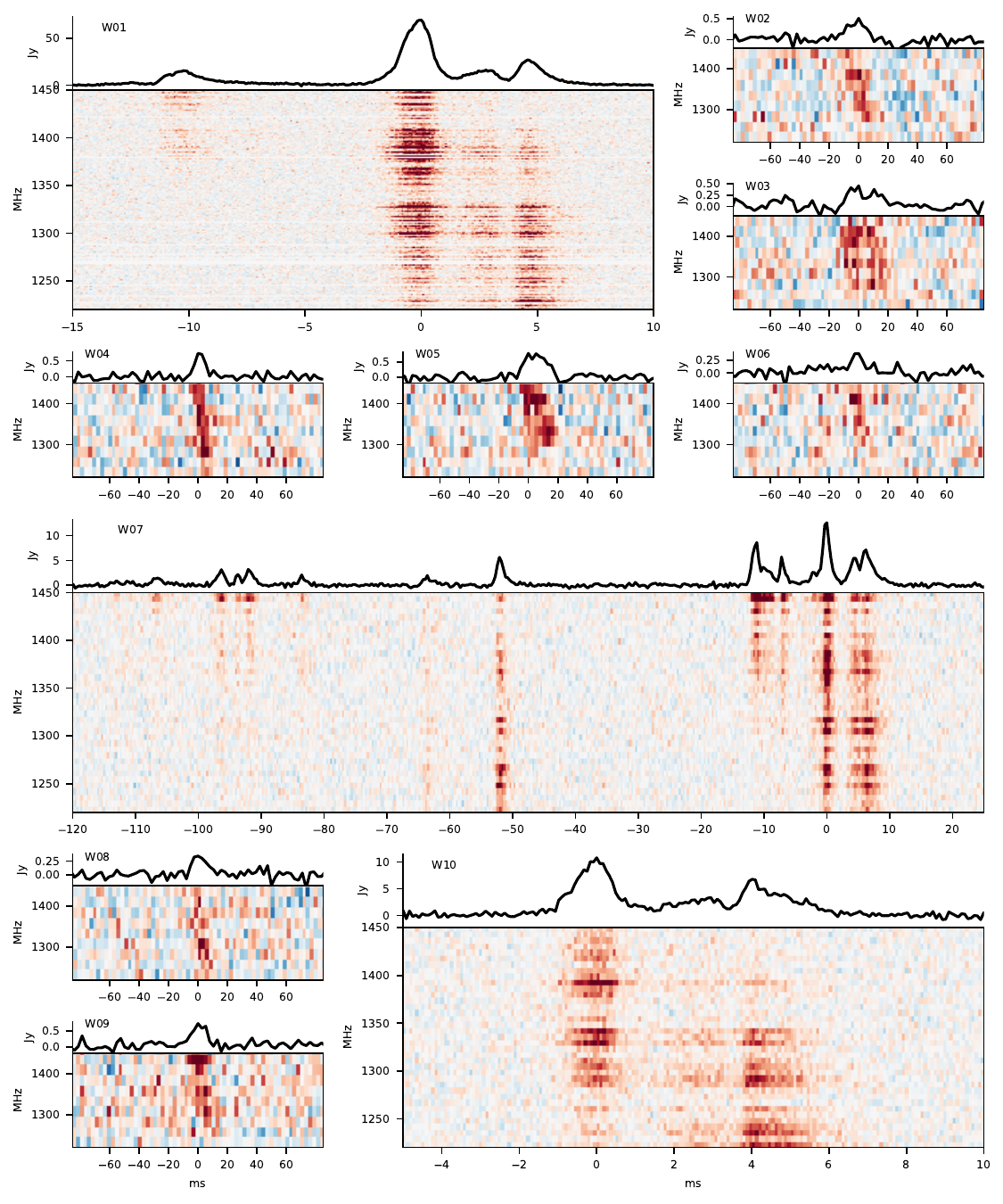}
 \caption{Spectra (bottom subpanels) and intensity profiles, integrated 
 over the frequency band (top subpanels), of the 
 ten bursts discovered within the ALERT survey. For plotting, the spectra 
 were normalized by the mean and the standard deviation in the off-burst 
 region in each sub-band, and the colors were saturated at $\pm10\sigma$ 
 for W01, W07, and W10, and at $\pm4\sigma$ for the other bursts. Bursts 
 W01 and W10 are plotted with the  original $t_\mathrm{res}=81.9\,\mu$s, burst 
 W07 with 0.33\,ms and the rest with 2.62\,ms. The frequency resolution is 
 25\,MHz for all bursts except W07 and W10 (6.25\,MHz), and W01 (0.48\,MHz).
 The time in ms is zeroed on the burst peak. Spectra at frequencies above 
 1450\,MHz are not shown, since the signal there is mostly corrupted by RFI.}
\label{fig:bursts}
\end{figure*}

Table~\ref{table:transientX} lists the times of arrival, best DM, integrated 
S/N and boxcar filter width for all 24 candidates.  The majority  (83\%) 
have DMs larger than 411\,\dmu, the DM of the brightest pulses from \src. 
However, \texttt{transientX} optimizes DM to maximize the S/N  and for 
some bright pulses this clearly compromises the intrinsic spectro-temporal 
structure.

About 70\% of all candidates come from one observing session, 2022 February\,01. 
The 2021 July sessions and session 2022 February\,06  yield two candidates each, 
and  session 2022 February\,05  resulted in only one (relatively bright) burst. 
It is possible that some of our faint candidates are due to chance noise 
fluctuations. In order to estimate the rate of occurrence of such noise 
candidates, we performed the same search in DM range between 660 and 760\,\dmu, 
leaving all other parameters intact. In this manner three candidates were 
visually filtered from about 500 candidates per session. Two candidates 
were detected in session 2021 July 04 and one in 2022 February 05, their 
integrated S/N values were $\leq8.1$, widths ranged from 0.3 to 6.3\,ms, and 
on the diagnostic plots the spectra looked indistinguishable from the 
spectra of faint FRB candidates from Table~\ref{table:transientX}. 

Table~\ref{table:transientX} lists 13 pulses with  $7<\mathrm{S/N}\leq8.1$, 
significantly more than the $\sim$3 that would have been expected by chance 
detections following the test described above. This likely means that some 
of those candidates were emitted by \src, 
although we cannot tell which ones exactly. Their faintness precludes 
any meaningful analysis, thus we do not further include them in the sample.
The spectra of the remaining ten bursts were next computed using \texttt{dspsr} 
and are shown in Fig.~\ref{fig:bursts}.

Calibration of the ALERT FRBs is performed with the help of drift scan 
observations of the bright quasars 3C147, 3C286, and 3C48 at the beginning 
and the end of each observing run \citep[cf.][]{Connor2020,Pastor-Marazuela2022a}.
We used the calibrator observation  closest to a given \src\ session and 
estimated the System Equivalent Flux Density (SEFD) using the known quasar 
flux \citep{Perley2017}. For 2022 February 01, the SEFD was 94\,Jy and for 
the last two sessions it was 82\,Jy. There is a slight (10\%)
variation within the band which was ignored. However, we took into account 
the excised parts of the band. Scaling the SEFD according to the radiometer equation
resulted in about
170 or 200 times smaller SEFD for the band-integrated signal at the original time
resolution (the denominator in 
Eq.~A1.16 of \citealt{Lorimer2005}), 
depending on the number of excised channels.
Following \citet{Pastor-Marazuela2021} we assume 20\% errors on the flux density values.

The quasar driftscan observations in the end of the 2022 June-July observing 
run failed, but test observations of pulsars immediately before and after \src\ 
observations did not indicate any malfunction. Taking a typical SEFD 
of 85\,Jy as derived in \citet{vanLeeuwen2022}, the fluence  
$F =\mathrm{S/N}\times\mathrm{SEFD}\sqrt{w_\mathrm{sec}}/\sqrt{n_\mathrm{pol}\mathrm{BW}}$
of the faint (i.e., below the adopted S/N threshold) bursts from 
Table~\ref{table:transientX} ranges from 0.7 to 4\,Jy\,ms, 
which is comparable to the limits by \citet{Mao2022} 
and larger than the 0.02\,Jy\,ms limits for 5 ms pulses after 
the emission quenching as reported by \citet{Xu2021}. Still,
we believe that the small excess of burst candidates detected 
near the plausible source DM (four around 410\,\dmu\ versus one 
candidate around the incorrect DM of 610\,{\dmu})
does not provide compelling evidence for the detection of faint FRBs 
between \Spring and \Fall {activity window}s. More robust estimates 
of the chance probabilities of such detection are beyond the scope of 
this work.

Among the ALERT bursts, W07 stands out because of its complex 
structure, appearing to consist of two groups of pulses with separations 
comparable to the duration of the groups themselves, clearly visible 
in Fig.~\ref{fig:bursts}. The burst was actually detected as two 
separate events by  \texttt{transientX}, but in what follows we   
analyze it as one cluster-burst, following the convention of 
\citet{Zhou2022}, who, based on the waiting time distribution of 
the emission peaks from \citet{Xu2021}, define such a  ``cluster-burst'' 
as a collection of emission peaks with a separation less than 400\,ms, 
without signs of bridge emission between them. 

The ALERT rate of 3 bursts per hour is seemingly smaller than the 
5.6--45.8 hr$^{-1}$ reported by \citet{Xu2021}. However, taking into 
account only those FAST pulses which satisfy the width-dependent fluence 
threshold based on an integrated ALERT S/N of 8.1, 
$F=0.94\sqrt{w_\mathrm{ms}}$\,Jy\,ms, we find that the FAST rate 
was close to the ALERT values in the beginning of the FAST observing 
campaign and extrapolates to 10--14 bursts per hour for the FAST 
sensitivity limits. 

On 2022 February\,01, no pulses other than W07 appear clustered.
Session 2022 February 05  yielded only one relatively bright burst, 
11 minutes into the observation, but no other bursts, even faint ones, 
were detected later. \citet{Xu2021} report at least one instance of 
a change in rate by a factor 3 (5 on their full sample) between daily 
sessions. The absence of emission after 2022 February 5 is unlikely 
to be the end of this {activity window}, since \citet{Takefuji2022} observed 
a burst at 2.3\,GHz on 2022 February 18. After that no other detections 
were reported.

\subsection{Simultaneous observations by GMRT and ALERT}
\label{sect:simult}

Apertif and the GMRT were co-pointing on 2021 February 05 from 18:20--19:30 UT, 
% [59615.76388889, 59615.8125    ]
and on February 06 from 16:30--19:30 UT.  % [59616.6875, 59616.8125]
During this time there is one burst detection, G07 (Table~\ref{table:FRBs_GMRT}).
There is no evidence for this same burst in the Apertif data.
We conclude the burst emission is band limited, and does not extend from 650\,MHz 
up to 1.4\,GHz, similar to the behavior we have found earlier in 
\srcRiii\ \citep{Pastor-Marazuela2021}.

%%%%%%%%%%%%%%%%%%%%%   Burst analysis
\section{Burst analysis}\label{sect:bursts}

\begin{table}
\begin{center} 
\caption{Observational properties of detected FRBs.}
\begin{tabular}{cccccc} 
\hline\\ %[0.01cm]
\parbox{1.0cm}{\centering  Burst \#}&
\parbox{1.4cm}{\centering  DM (\dmu)}&
\parbox{1.0cm}{\centering Equiv. width (ms)} &
\parbox{1.0cm}{\centering Span (ms)} &
\parbox{1.0cm}{\centering Peak Flux (Jy)} &
\parbox{1.0cm}{\centering Fluence (Jy\,ms)} 
\\ [0.3cm]
\hline\\
%# DM DMer w_boxcar  w_acf range speak fluence
W01 & 410.06(13) & 3    &   15  & 69.6 & 215.3 \\
W02 &            & 10  &       & \phantom{0}0.5  & \phantom{00}5.1    \\
W03 &            & 15  &       & \phantom{0}0.5  & \phantom{00}7.7 \\
W04 &            & 9  &       & \phantom{0}0.7  & \phantom{00}6.2 \\
W05 &            & 15  &       & \phantom{0}0.8  & \phantom{0}12.2 \\
W06 &            & 18  &       & \phantom{0}0.4  & \phantom{00}7.2 \\
W07 & 410.01(34) & 7    & 112   & 12.6 & \phantom{0}98.6 \\
W08 &            & 14  &       & \phantom{0}0.3  & \phantom{00}4.2   \\
W09 &            & 15  &       & \phantom{0}0.7  & \phantom{0}10.9  \\
W10 & 409.89(42) &  3   &   5   & 10.8 & \phantom{0}28.5 \\
\\[0.1cm]
\hline 
\end{tabular} 
\tablefoot{ The columns are: burst number, 
DM as measured with \texttt{DM\_phase}, equivalent width from \texttt{transientX}, 
span (distance between peaks of the first and the last sub-burst), peak flux density, 
and fluence. \label{table:FRBprops}}
\end{center}
\end{table}

\subsection{Dispersion measure} \label{sec:DM}

%%%%%%%% Fig DM vs SN
\begin{figure}
\centering
\includegraphics[width=0.4\textwidth]{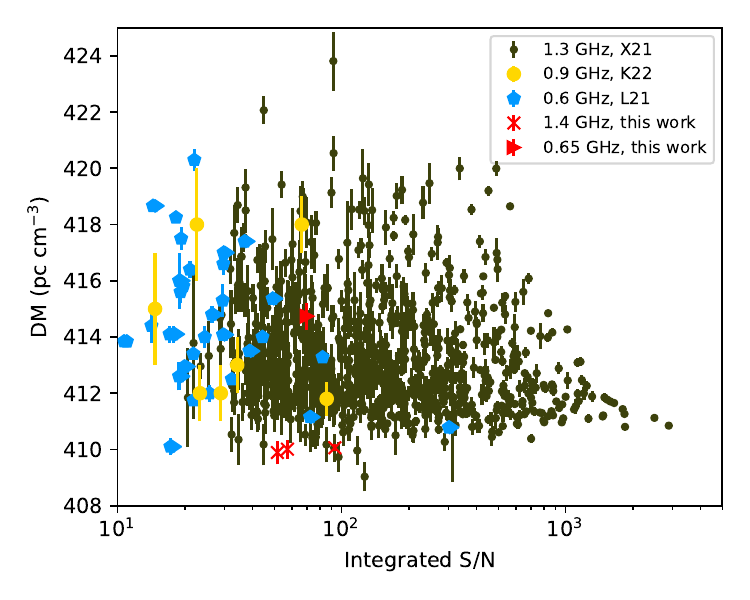}
\caption{Compilation of DM measurements from previous studies: \citet[][X21]{Xu2021}, 
\citet[][K22]{Kumar2022}, \citet[][L22]{Lanman2022}, and this work, plotted as a 
function of integrated S/N. }
\label{fig:DM}
\end{figure}

%%%%%%%% Fig DM phase
\begin{figure*}
\centering
\includegraphics[width=0.25\textwidth]{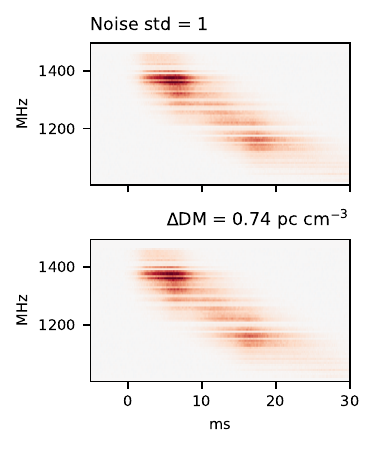}\includegraphics[width=0.25\textwidth]{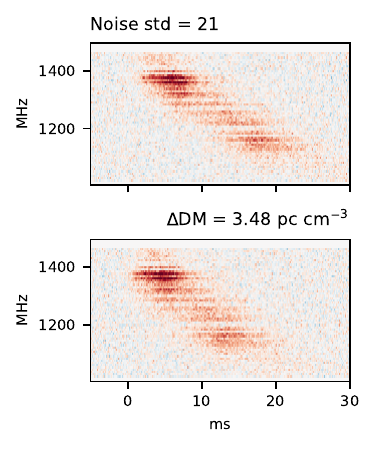}\includegraphics[width=0.25\textwidth]{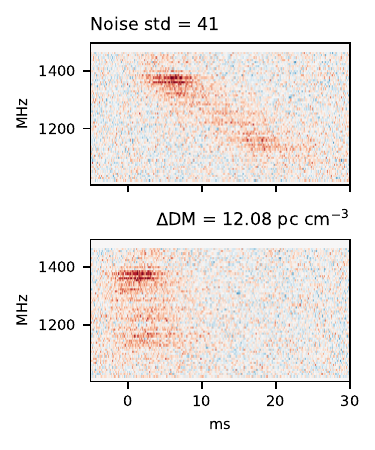}
\includegraphics[width=0.25\textwidth]{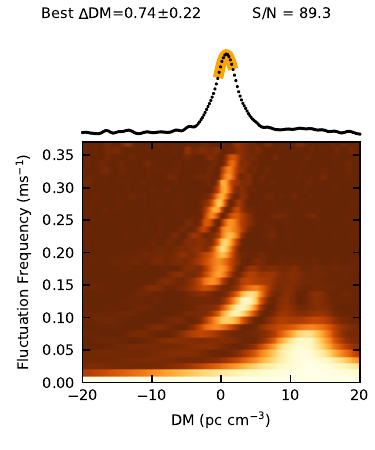}\includegraphics[width=0.25\textwidth]{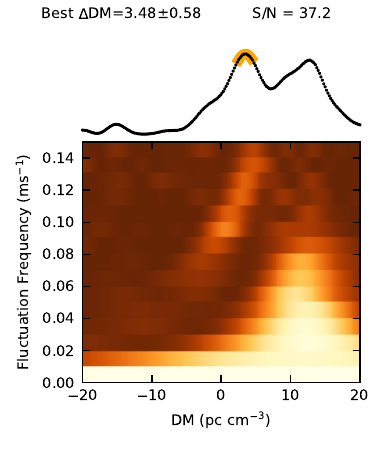}\includegraphics[width=0.25\textwidth]{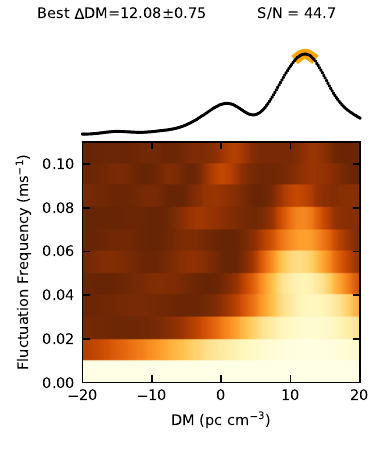}
\includegraphics[width=0.75\textwidth]{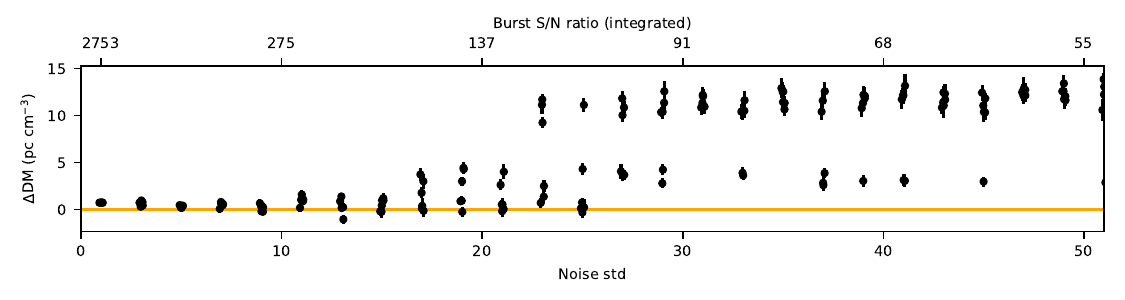}
\caption{Influence of noise on the DM estimate from $\texttt{DM\_phase}$. 
Burst \#377 from Fig.~6 in the extended data in \citet{Xu2021} was normalized 
in the off-burst window in each channel (such that the standard deviation 
of the noise $\mathrm{std}=1$), and Gaussian noise was added in each channel. 
Each simulation was then run through $\texttt{DM\_phase}$ with an automatic 
spectrum cut. \textit{Top row:} Example of spectra before and after DM 
correction for three different values of noise std. 
\textit{Middle row:} Corresponding \texttt{DM\_phase} spectra. \textit{Bottom row:} 
$\Delta\mathrm{DM}$   vs a set of noise realizations, six simulation runs per 
each noise std. For clarity, some jitter along the horizontal axis was added 
to each set of points.  }
\label{fig:DMphase}
\end{figure*}

Dispersion is an important characteristic of FRBs as it measures 
the integrated electron density on the line of sight (LOS)
provided that  the effect of dispersion can be separated from any
intrinsic spectral shape of the burst. Variations of 
the DM may indicate a complex and dynamic circum-burst environment, since rapid 
DM changes are not expected at the galactic and inter-galactic level.
When combined with the RM, the DM places limits on the magnetic field 
strengths encountered by the bursts \citep[e.g.,][]{Xu2021, Lu2023}.

As a quantity, the DM characterizes the magnitude of the pulse delay 
-- a delay that is inversely dependent on the square of the observing 
frequency. In the absence of any a priori knowledge about the intrinsic 
spectral shape of a burst, the DM can be estimated by maximizing the 
coherent power in the burst, across the observing bandwidth 
\citep{Seymour2019}. This technique is  employed by 
\texttt{DM\_phase},\footnote{\url{https://www.github.com/DanieleMichilli/DM_phase}} 
which operates as follows. During dedispersion, the time series from 
each particular frequency channel is shifted to counteract the expected 
dispersive delay for this frequency. For a Fourier transform along the time
axis, this operation corresponds to a multiplication by $\exp(i\phi)$, 
with $\phi = \omega\,C\,\mathrm{DM}/\nu^2$, where $\omega$ signifies 
the Fourier frequency. The phase of the Fourier transform $\phi$ is 
integrated over $\omega$ and $\nu$, and the resulting dependence of 
integrated coherent power on the trial DM values is examined for peaks. 
The error of the DM measurement is assumed to be the error of peak 
position determination. 

We have utilized \texttt{DM\_phase} with an automatic cutoff along the 
$\omega$ axis to measure the DM for all our bursts that have an integrated 
$\mathrm{S/N} > 50$. The GMRT sample yielded one such burst, with a single 
component and slightly asymmetrical pulse shape. For this burst, the 
DM was measured to be $\mathrm{DM} = 414.73\pm0.48$\,\dmu. 

The WSRT sample contained three sufficiently bright bursts, all composed of a few 
distinct components. The DM measured here was lower, {410\,\dmu}, 
with a characteristic error of 0.3\,\dmu\ (Table~\ref{table:FRBprops}).
The initially reported measurement of \mbox{$\mathrm{DM} = 410.9\pm0.2$\,\dmu} 
for bursts W01 and W07b \citep{Atri2022} was based on the less precise 
method of straightening the pulse structures visually. Both the GMRT and 
WSRT measurements agree with the distribution of DMs within the respective 
{activity window}s \citep{Xu2021, Kirsten2024}.

The structure-maximizing method of DM estimation is a de-facto standard 
in the FRB field at the moment. However, it does not provide unambiguous measurements 
for all bursts. For some FRBs, optimal DMs determined from the upper/lower 
halves of the observing band are inconsistent with each other, and burst 
spectral features cannot be aligned across the whole band 
\citep[see, e.g.,][]{Platts2021, Kumar2022, Zhou2022}.

Below we  demonstrate that the existence of a DM value that maximizes 
burst coherent structure does not mean that this DM can be readily used 
to measure the integrated electron density along the path the emission traveled.
 Depending on the burst spectral shape, the structure-maximizing 
DM may be biased by fine structure buried in noise. We illustrate this 
using the comprehensive study of \src\  bursts  by \citet{Xu2021}.
The authors compiled a set of DM values measured with \texttt{DM\_phase} 
for bursts with integrated S/N  of 20--3000 and a variety of 
time-frequency profiles. The authors rule out secular DM trends on the 
level of 2.9\,\dmu\ per two months,  but record a large spread of DMs 
(with variations on the order of a few \dmu) within individual 2 hr sessions, 
sometimes on a timescale of less than a minute or, remarkably, even less 
than a second.

There is a correspondence between the reported DMs and the integrated 
S/Ns of the bursts in the \citeauthor{Xu2021} dataset: the DMs of 
the brightest bursts are listed as lower (around 411\,\dmu), whereas 
the reported DMs of fainter bursts are spread between 409 and 
424\,\dmu\ (Fig.~\ref{fig:DM}). \citet{Kumar2022} and \citet{Lanman2022} 
report a similar spread of DM values for comparable S/N values.\footnote{
While \citet{Kumar2022} uses \texttt{DM\_phase}, \citet{Lanman2022} 
use \texttt{fitburst}, which models bursts as a collection of Gaussian 
components with frequency-dependent amplitudes convolved with a 
scattering function (see \citealt{CHIMECat2021}, for a discussion on the 
limitations of this approach).}

The questions are whether this correlation signifies a causative relationship with the intrinsic 
brightness of the burst and/or whether it is related to the method or to 
instrument noise. To investigate the effect of noise on the performance 
of \texttt{DM\_phase}, we utilized a nine-burst sample of burst spectra that were 
made public by \citeauthor{Xu2021} at the time of manuscript preparation.\footnote{
After our analysis was performed, an extended version of burst spectra sample
from \citet{Xu2021} have been made public \citep{Wang2023a}.}
We  take burst \#377 from \citet{Xu2021} as an example, 
and discuss other bursts afterward. 

The observed behavior of burst \#377 is shown  in the top 
left panel of Fig.~\ref{fig:DMphase}.
It has three partly merged components and an overall drift in frequency.
The spectrum of the $\omega$-resolved power versus the trial DM shows 
different zones corresponding to these scales. If we now add Gaussian 
noise to each frequency channel of the data (pre-normalized by the 
standard deviation of noise in the off-pulse region), the fine 
temporal structure is washed out ever more, and above a certain amount 
of added noise only the low-$\omega$ feature remains. Above this edge 
in additional noise (i.e., below this boundary in terms of S/N),
$\texttt{DM\_phase}$ aligns the entire spectrum, converging on a 
DM about 10\,\dmu\ higher that the original value. The step is clearly 
visible when comparing  the middle column in Fig.~\ref{fig:DMphase} to 
the right column. 

We note that even without such additional noise the DM derived with 
\texttt{DM\_phase} depends on the choice of the frequency and time 
averaging, as well as the channel normalization method. The resulting 
spread of DM measurements is a  few times larger than the estimated 
\texttt{DM\_phase} uncertainties. The same level of discrepancy is 
observed between DMs from \citet{Xu2021} and our measurements. Also, 
on our normalized data we measure integrated S/N of about 1.5 times 
larger than reported by \citet{Xu2021}. This could be partly 
attributed to normalization or, partly, be due to a finer grid of trial 
widths we used for the integrated S/N calculation. We also note that
the some of the information in the header of burst spectra 
that accompany \citet{Xu2021} (e.g., DM and cardinal burst number) 
deviates from the corresponding entries in their data table. 

How much the DM varies with pulse S/N, as well as the character of 
this variation (e.g., with or without a step), is determined by 
the  temporal-spectral shape of the burst. The least amount of variation, 
less than a few times the \texttt{DM\_phase}-reported error, was 
recorded for bursts with distinct, widely separated components 
(e.g., bursts \#779, \#1377, and \#1398 from  Extended Data 
Figure~6 in \citealt{Xu2021}). Mean while,  bursts with an overall 
drift in frequency (e.g., \#377, \#460) exhibited steps of 
$1-10$\,\dmu\ even at integrated S/Ns as large as  100. Thus, for the 
majority of the burst population in Fig.~\ref{fig:DM}
the DMs are likely overestimated if unresolved 
drift in frequency was present. 

FAST provides by far the largest and brightest sample of bursts, 
since for other telescopes the S/N is usually smaller, meaning that 
DM overestimation is  widespread. As this is a matter of S/N, the 
same effect will occur for very bright bursts observed with 
  less sensitive telescopes. 
{It is worth noting that a DM measurement bias introduced by  drifting of separate burst components (that
  can go unrecognized if the burst does not have sharp sub-bursts) was investigated on a large sample of bursts from
  another prolific FRB repeater, {\srcRi}, by \citet{Jahns2023}. In that work, the measured tilt of each individual
  sub-burst  was directly converted to the resulting change in the fitted DM. This equivalent DM can be used as an upper
  limit on structure-maximized DM bias, since the algorithm for the latter is typically using information from a
  collection of sub-bursts, which can have different drift rates. The upper limits on this DM bias have the same range as in our work, namely $\leq 12$\,\dmu. While the authors speculate that the actual bias in structure-maximized DM estimates is likely to be less than 1\,{\dmu}, no direct studies have been performed. }

Assuming that the DM excess of \citeauthor{Xu2021} is indeed due 
to absorption of the trombone effect, we can calculate the absorbed 
drift rate from an extra dispersion delay between the edges of 
the band. For a $\mathrm{DM} \geq 412$\,\dmu, the frequency drifts 
are between 25 and 225 MHz/ms, comparable to the values found 
 by \citet{Zhou2022}, who determine 
their DM values from a sample of bursts with sharp, separate components.

\begin{figure}
\centering
 \includegraphics[width=0.4\textwidth]{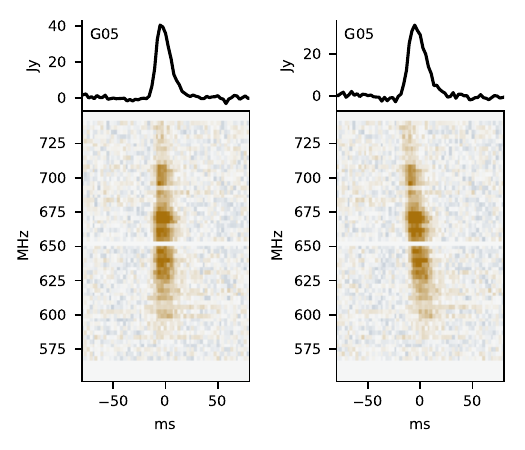}
   \caption{Burst G05 dedispersed with structure-maximizing 
   $\mathrm{DM}=414.73$\,\dmu (\textit{left}) 
   and with $\mathrm{DM}=411$\,\dmu (\textit{right}), 
   suggested by the brightest bursts observed by FAST.}
\label{fig:DM_G05}
\end{figure}

If the true DM at the time of the GMRT observations was 411\,\dmu, 
then the DM measured from burst G05 would imply a drift rate of 8\,MHz/ms
(Fig.~\ref{fig:DM_G05}). It is useful to compare this value to the 
study of \citet{Marthi2022}, who calculated drift rates for a  sample 
of 48 FRBs recorded during a single session with the GMRT. In that 
study, a DM of $410.78\pm 0.54$\,\dmu\ was determined using a 
singular-value decomposition of the spectrum of the brightest burst, 
exhibiting a  sharp burst rise and a few partially merged components. 
All bursts in the study of \citet{Marthi2022} displayed trombone drift, 
with rates between 0.75 and 20\,MHz/ms. We thus conclude that the 
DM determined for G05 has some drift absorbed in it.

%%%%%%%% G05 with two DMs

For the \Fall\ {activity window}, \citet{Zhou2022} measure an average 
DM of  412.4(3)--411.6(3)\,\dmu, using only those pulses that show 
sharp edges or well-separated components. Individual measurements 
are spread within $\pm2$\,\dmu, larger than the reported errors. 
\citet{Kirsten2024} measure DMs consistent with \citet{Zhou2022} for 
that day. Statistically, their DMs for the \Winter\ {activity window} are 
no different from our measurements. Overall we conclude that reliably 
detecting any secular DM trend on the level of 1\,\dmu\ between 
{activity window}s requires a more robust method of dealing with the 
influence of burst structure than is currently used.

%%%%%%%% ACF
\begin{figure}
\centering
 \includegraphics[width=0.4\textwidth]{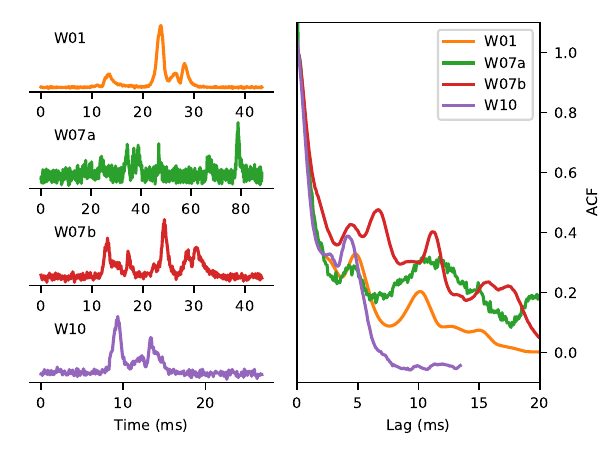}
   \caption{\textit{Left column}: Bursts W01, W07a, W07b, and W10 
   with potential quasiperiodic components. \textit{Right:} 
   Autocorrelation function from all four bursts.  }
\label{fig:ACF}
\end{figure}

%%%%%%%% Fig quasi-period.
\begin{figure}
\centering
 \includegraphics[width=0.5\textwidth]{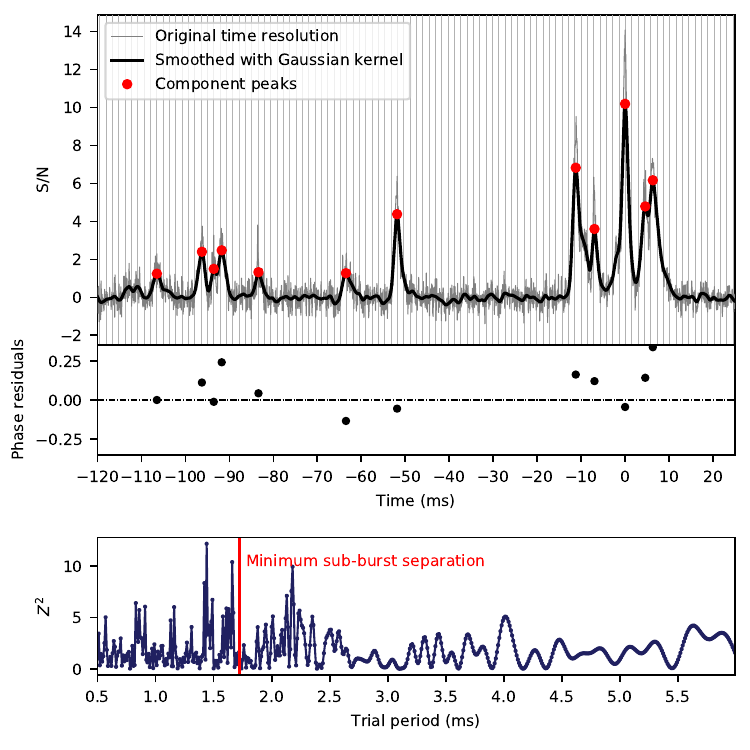}
   \caption{\textit{Top}: TOAs of individual sub-bursts of FRB 
   W07 (red dots) identified as peaks of the signal smoothed 
   with Gaussian kernel (thick line). The vertical lines show 
   periodicity with best quasi-period $P=1.44$\,ms. \textit{Middle:} Phase 
   residuals with respect to model based on the best-fit quasi-period $P$. 
   \textit{Bottom:} $Z^2$ statistic on the range of trial periods.  }
\label{fig:QPO}
\end{figure}

\subsection{(Quasi-)periodicity}

Some FRBs consist of multiple components arranged in a seemingly regular fashion. 
A reliable detection of such \mbox{(quasi-)}periodicity may have interesting
implications for FRB emission theories. For example, such periodicity may be 
a direct manifestation of the relatively fast spin period of an emitting 
compact object. 
Or, it may reflect the features of spark generation or non-stationary plasma 
flow in the neutron star magnetosphere \citep[e.g.,][]{Mitra2015}. Finally, 
quasi-periodicity of subpulse components appears naturally when FRB generation 
is driven by crust motion \citep{Wadiasingh2020b}.

So far, the most reliable detection of periodicity comes from a one-off FRB. 
\citet{CHIME2021} detected a 217\,ms periodicity in the nine components of 
{\FRBperiodic}. This  $6.5\sigma$ periodicity is consistent with beamed emission 
from a neutron star rotating at $\sim$5\,Hz. Other detections are close to the 
$3\sigma$ level and exhibit quasi-periods ranging from submilliseconds to tens 
of milliseconds \citep{Pastor-Marazuela2022a, CHIME2021}. 

\src\ presents an interesting example of a repeating FRB source characterized 
by apparent periodicity in some of its sub-bursts. Nevertheless, timing 
analysis of the sub-burst TOAs derived from 53 bursts within 
the \Fall\ activity window, as conducted by \citet{Niu2022}, did not reveal a 
single period with possible harmonics, but unveiled a broad distribution of 
periods spanning the range of 1 to 10\,ms, with an extended tail reaching 
up to 50\,ms. All identified periods exhibited significances not exceeding 
$3.9\sigma$ of the normal distribution.

In our sample, only bursts W01, W07, and W10 exhibit multiple components with 
potentially periodic spacing. To test for this subsecond periodicity we computed 
the auto-correlation  function (ACF) of W01, W10 and the two parts of W07 
(Fig.~\ref{fig:ACF}), but no prominent periodicity was found. The quasi-periods 
corresponding to the peaks on ACF are in good agreement with the distribution 
obtained by \citet{Niu2022} on a larger sample of bursts.

We complemented the ACFs with a timing analysis, which may be more sensitive to short 
periodicities potentially buried in the zero-lag peak of the ACF. In the
timing analysis, periodicity is searched for in the sample of sub-burst TOAs, 
and the accuracy of the timing analysis is greatly influenced by the precision 
of the TOA measurement. If sub-bursts have complex shape and are closely spaced, 
it becomes difficult to determine which parts of the time series belong to 
different sub-bursts, and which reflect the intrinsic shape of an individual 
sub-burst. An example of this can be seen in the sub-bursts of FRB W07 on the 
upper subplot of Fig.~\ref{fig:QPO} around $t=-10$\,ms. 

For the subsequent analysis, we defined each TOA as the time stamp of the peak on 
the signal smoothed with a Gaussian kernel with a five-sample standard deviation. 
The peaks were located using \texttt{scipy.signal.find\_peaks}.

For bursts with prominent, ostensible quasi-periodicity it is logical to assume 
that the period is close to sub-burst separation or its integer 
multiplicative (e.g., FRB 20191221A in \citealt{CHIME2021}, where sub-bursts 
arrive every one or two periods). The situation becomes more complicated 
when bursts are less frequent, as in \src. Searching for periods 
around  the minimum sub-burst component separation  then leads to a 
large variation of these periods from one burst to another, as was 
demonstrated by \citet{Niu2022}.

In what follows we thus do not  assume that the minimum separation 
between sub-bursts strictly constrains the possible periodicity. 
Instead, we performed a uniform search over a grid of trial periods 
using the $Z^2$ statistic:
\begin{equation}
Z^2 = \frac{2}{N}\left[ \left(\sum_{i=0}^{N-1} \cos \phi_i \right)^2  + \left(\sum_{i=0}^{N-1} \sin \phi_i \right)^2\right].
\end{equation}
Here $\phi_i = 2\pi t_i/P$ is the phase of \textit{i}th sub-burst component 
computed with trial period $P$. The statistic $Z^2$ was maximized over the range of $P$ between 
0.5 and 6\,ms with an increment of 0.01\,ms. We estimate the significance by
repeating the analysis on $10^5$ samples of random TOAs, keeping  the first and last 
$t_i$ fixed, and drawing the remaining TOAs from a  uniform distribution
spanning $(t_0,\,t_{N-1})$. 

%%%%%%%% Fluence distribution plot
\begin{figure*}
\centering
   \includegraphics[width=\textwidth]{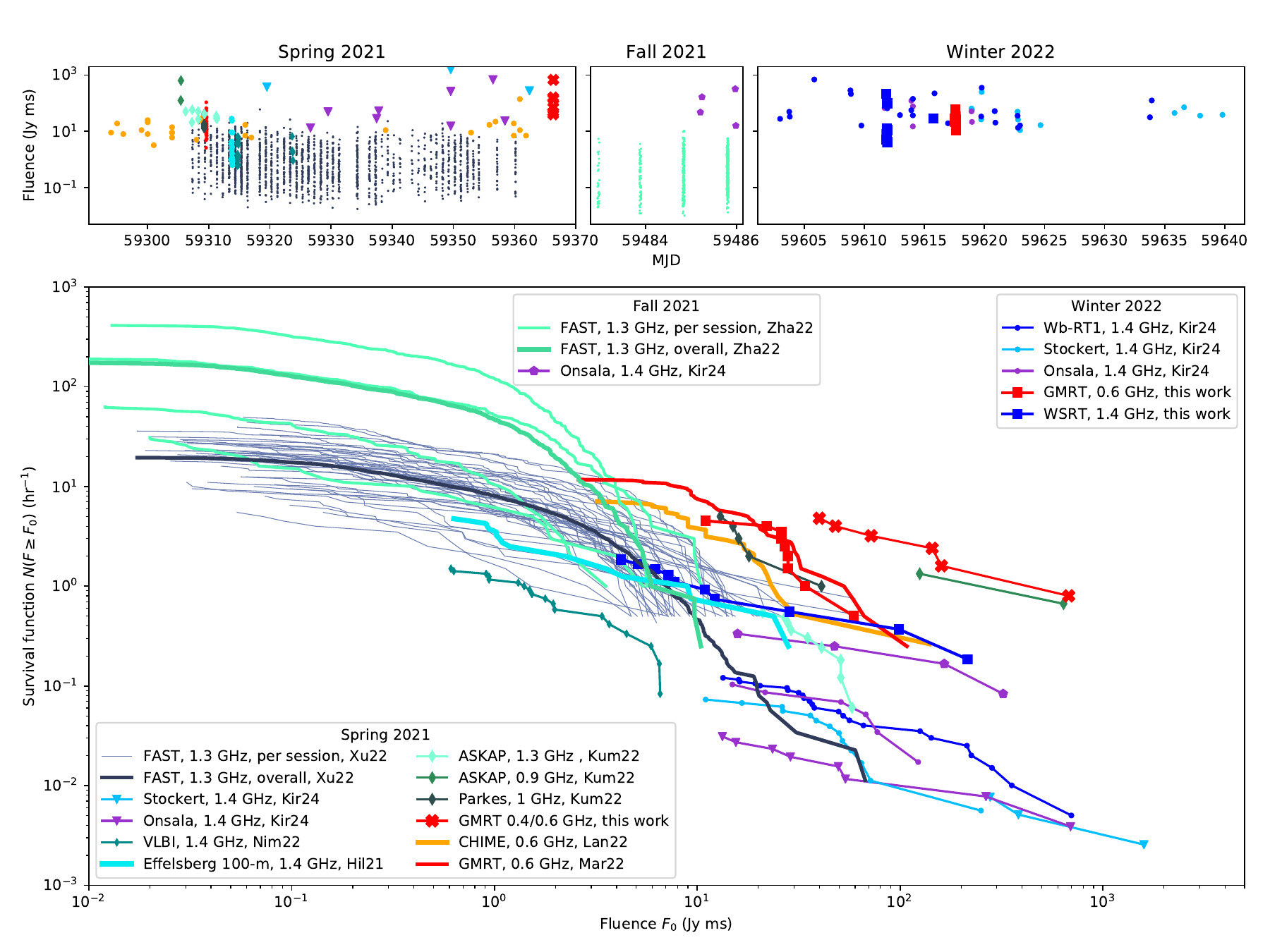}
    \caption{\textit{Upper row:} FRB fluence vs MJD of arrival for three 
    {activity window}s. The markers and colors correspond to the legend in the 
    \textit{lower} panel (see inset), which shows rate survival function vs 
    burst fluence. FRB information was collected from \citet[][Lan22]{Lanman2022},  
    \citet[][Xu21]{Xu2021}, \citet[][Nim22]{Nimmo2022}, 
    \citet[][Hil21]{Hilmarsson2021}, \citet[][Kum22]{Kumar2022}, 
    \citet[][Mar22]{Marthi2022}, \citet[][Zha22]{Zhang2022}, and
    \citet[][Kir24]{Kirsten2024}. Where possible, fluence distributions 
    obtained from observations on different days and with 
    different telescopes were computed separately}.
\label{fig:fluence_distr}
\end{figure*}

Applying this analysis to the 11-component burst from \citet{Niu2022}, we find an optimal 
$P$ of 3.07\,ms, $Z^2=11.48$. In the batch simulated of sets containing 11 random TOAs, 
99.75\% of sets had a maximum $Z^2<11.48$ for $P=3.07$\,ms. This equates to a
significance for the observed burst equivalent to $2.84\sigma$ for a normal distribution.
The results are 
close to those obtained by \citeauthor{Niu2022}: $P=3.06$\,ms with significance of 
$3.3\sigma$.\footnote{Differences may stem from inaccuracies in determining TOAs 
from the digitized version of their Fig.~7.} However, since we searched over 
a grid of $P$, the number of trials should be taken into account. Comparing the maximum $Z^2$ 
score of the real data to the pool of maximum $Z^2$ scores lowers the significance 
to $0.3\sigma$. Thus, for these weak signals the presence of a priori 
constraints on $P$ is crucial for obtaining a  significant result.

Among the bursts in our sample, W07 exhibited {the strongest evidence for} quasi-periodicity{. This
  candidate had} a period of $1.44$\,ms, and a significance of $3.3\sigma$ directly, and $0.68\sigma$ after correction 
for the number of trials (Fig.~\ref{fig:QPO}). For Bursts W01, W10, and the two sub-burst groups of W07, 
no discernible periodicity with single-trial significances greater than $2\sigma$ was identified. 
It appears the sub-bursts lack prominent quasi-periodicity, at least when not considering 
sub-burst shape appropriately or in the absence of  motivated constraints on $P$.

Under the assumption that sub-bursts  appear almost every period, and that 
$P$ can change from one burst to another, we examined the distribution of $P$ from 
\citet{Niu2022} in order to investigate whether the frequencies  $1/P$  were clustered around 
specific harmonics, as predicted by the crust motion and low-twist theory \citep{Wadiasingh2020b}. 
We did this by obtaining the eigenmode $l$-number from the formula 
$\nu = 0.5 \nu_0\sqrt{(l-1)(l+2)}$ over a  range of $\nu_0$ trials.
For each trial $\nu_0$ we examined the deviations of the obtained values for $l$ from 
their respective nearest integer counterparts. These distributions appeared to be 
uniform and indistinguishable from the same distributions computed on sets of 
uniformly distributed random period values. 
Although no clustering was hence found, we cannot disprove crust motion  
 theory this way. Since the eigenfrequency depends also on the strength of 
magnetic field at the location and time of the crust motion, it may vary from 
one burst to another \citep{Wadiasingh2020b}.

\subsection{Burst fluences}

%############ Figure Kirsten-Xu 
\begin{figure}
    \includegraphics[width=0.5\textwidth]{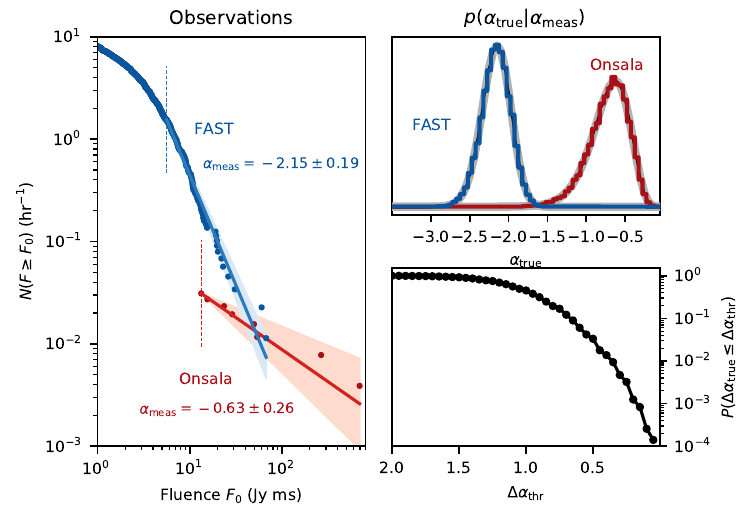}
\caption{
\textit{Left:} Survival functions of the fluence distributions from the FAST and Onsala 
observations during the \Spring\ {activity window}, together with MLE PL fits. The vertical 
lines mark the optimal minimum fluence as found by the  \texttt{powerlaw} 
package, after modification according to 
Eq.~\ref{eq:a_unbiased}-\ref{eq:a_err_unbiased} in Appendix~\ref{app:PL_fit}. 
\textit{Top right:} Posterior distributions of $\at|\am$. 
The gray lines indicate gamma distributions from Eq.~\ref{eq:post_prob}. 
\textit{Bottom right:} Probability of $\at(\mathrm{Onsala}) - \at(\mathrm{FAST})$ 
being smaller than the threshold value.}
\label{fig:Kirsten_distr}
\end{figure}

Tables~\ref{table:FRBs_GMRT} and \ref{table:FRBprops} list the peak 
flux densities and fluences of the bursts recorded in this study. 
Prior to measuring the peak flux densities, the band-integrated 
time series were averaged by several time bins (see the caption of 
Fig.~\ref{fig:bursts}). For fainter bursts, the peak flux density 
depends on how much of such averaging was performed.

The fluence was estimated by summing the signal in a fixed-size 
window around the burst peak. For all bursts except W07, this window 
size comprised 60\,ms. For W07 we used a larger window, from  
$-120$ to 20\,ms, to encompass all components.
The fluence of W01 is in agreement with the CB17 detection 
we reported in \citet{Atri2022}. For W07, we measure a 3 times
larger fluence in the larger window, as we are now including the 
component group W07a, in contrast to the earlier reported results 
from the standard pipeline.

\subsubsection{Fluence distributions}

Statistical distributions of burst fluences provide valuable information 
about the FRB emission mechanism, and about propagation in the circum-burst 
environment \citep{Cui2021,Xiao2024}.
Owing to the large number of bursts observed, \src\ has the potential 
to offer one of the best-measured distributions of burst fluences among 
repeating FRB sources. In practice, however, the distributions from 
different observations exhibit little agreement with each other 
(Fig.~\ref{fig:fluence_distr}). The apparent mismatch is not entirely
caused by the difference between the mean event rates at different epochs,
but the shape of the distributions themselves seem to have intrinsic
changes. 

A series of dedicated observations by the FAST telescope demonstrated 
that the \src\ burst rate is highly variable (by up to two orders of 
magnitude), both within and between {activity window}s \citep{Xu2021, Zhang2022}. 
The fluence distribution is bimodal \citep{Zhang2022}, but unfortunately, 
the pulse rate is insufficient for exploring any changes in the 
distribution shape on the timescale of the rate change, although some 
studies have been performed \citep{Sang2023}.

The fluences of bursts from WSRT observations fall within the range of 
fluences reported by other authors. Fainter bursts likely belong to the 
fainter component of the bimodal fluence distribution seen in the  FAST observations 
\citet{Zhang2022}. No contemporaneous observations were conducted 
during \Winter\ activity window, however, and the position of the fainter component 
is known to shift from one activity window to another. Fluence distributions for 
\Winter\ activity window recorded with WSRT and \Spring\ activity window recorded
with GMRT are flatter than normal, although similarly
flat distributions have been recorded previously.

\subsubsection{Fitting methods}
\label{subsec:fluence_fit}

Generally, a cumulative distribution of burst fluences can be approximated with a
power-law (PL) distribution, with a possible flattening at lower fluences, either 
intrinsic or due to sample incompleteness close to observational sensitivity limit. 
PL fits are easy to perform, have only few free parameters, and allow for quick 
comparisons with theoretical models \citep[e.g.,][]{Wadiasingh2020a}. There are, 
however, limitations that should be kept in mind as these can bias the physical 
interpretation of the results.

One of the caveats concerns the fitting method. Historically, fluence 
distributions of individual pulses of pulsar radio emission have been 
approximated with PL functions by performing a least-squares linear fit 
to the survival function on a log-log scale (hereafter the ``graphical method''), 
and this practice is still sometimes used for the FRB fluence distribution 
\citep{Popov2007,Bilous2022,Pastor-Marazuela2021,Kirsten2024}. 
Despite the ostensible transparency of the graphical fitting method, the 
least-squares minimization does not provide an accurate and unbiased estimate of 
the PL parameters -- even if the number of the sample is quite large by FRB standards 
\citep[$\sim100$ pulses, see][]{Goldstein2004,Hoogenboom2006}. 
A Maximum Likelihood Estimator  (MLE) provides a much more accurate estimate 
of the PL index. In Appendix~\ref{app:PL_fit} we provide a detailed comparison 
of the two methods. 

%%%%%%%%%%%%%%%% Random subsample from distribution with flattening
\begin{figure}
\includegraphics[width=0.45\textwidth]{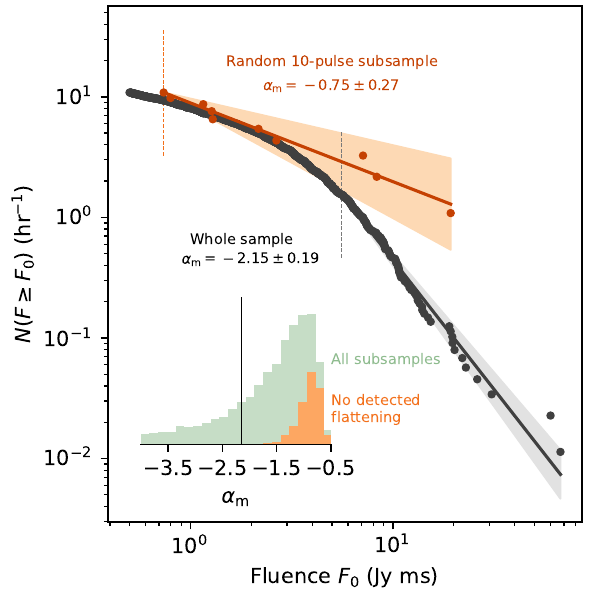}
\caption{Survival functions of the fluence distribution from the
\Spring\ FAST bursts \citep{Xu2021} 
{with fluences larger than 0.5}\,Jy\,ms, together with an example 
survival function from a random ten-pulse subsample. The vertical lines 
indicate the optimal minimum fluence values from the PL fit. The example 
survival function does not show apparent flattening. \textit{Inset:} 
Distribution of $\am$ for all 10$^{4}$ ten-pulse subsamples (green) 
and those that do not show flattening (orange).}
\label{fig:flattening}   
\end{figure}

%%%%%%%% Table with PLI fits
\begin{table*}[h]
\begin{center} 
\caption{Power-law indices for the survival functions of burst fluence 
distributions,   from published studies and from our measurements. }
\begin{tabular}{p{3.5cm}p{1.2cm}p{1.2cm}p{1.2cm}p{1.2cm}p{0.7cm}p{1.0cm}p{1.0cm}p{1.0cm}p{1.0cm}p{1.0cm}p{0.9cm}} 
\hline\\ %[0.01cm]
\parbox{3.5cm}{\centering Reference}&
\parbox{1.2cm}{\centering Telescope} &
\parbox{1.1cm}{\centering  $\nu_\mathrm{c}$\\MHz} &
\parbox{1.cm}{\centering MJD start} &
\parbox{0.7cm}{\centering $\am$ } &
\parbox{0.5cm}{\centering $\alpha_\mathrm{err}$} &
\parbox{1.0cm}{\centering min  $F_\mathrm{obs}$} &
\parbox{1.0cm}{\centering min  $F_\mathrm{PL}$} &
\parbox{1.0cm}{\centering max  $F_\mathrm{obs}$} &
\parbox{1.0cm}{\centering \# of FRBs} & 
\parbox{0.9cm}{\centering fraction PL} %fraction obeying PL} 
\\ [0.3cm]
\hline\\
%# DM DMer w_boxcar  w_acf range speak fluence
\multicolumn{11}{c}{Spring 2021} \\
\hline
\\[0.1cm]
\citet{Lanman2022} & CHIME & \phantom{0}600 & 59177 & $-1.22$ & 0.24 & \phantom{00}3.2\phantom{00} & \phantom{00}5.8\phantom{00} & 140\phantom{.000} & \phantom{00}33 & 0.88 \\
\citet{Kumar2022} & ASKAP & 1271 & 59306 & $-2.48$ & 1.01 & \phantom{0}21.0\phantom{00} & \phantom{0}27.0\phantom{00} & \phantom{0}58.0\phantom{00} & \phantom{000}9 & 0.89 \\
\citet{Xu2021} & FAST & 1250 & 59307 & $-2.15$ & 0.19 & \phantom{00}0.017 & \phantom{00}5.6\phantom{00} & \phantom{0}67.3\phantom{00} & 1715 & 0.08 \\
% For Parkes UWL I changed central freq of the band to central freq of FRBs since they were detected only in the lowest subband
\citet{Kumar2022} & Parkes & \phantom{0}850 & 59309 & $-2.19$ & 1.27 & \phantom{0}13.0\phantom{00} & \phantom{0}13.0\phantom{00} & \phantom{0}41.0\phantom{00} & \phantom{000}5 & 1.00 \\
\citet{Marthi2022} & uGMRT & \phantom{0}650 & 59310 & $-1.13$ & 0.18 & \phantom{00}2.6\phantom{00} & \phantom{00}7.2\phantom{00} & 108\phantom{.000} & \phantom{00}47 & 0.89 \\
\citet{Hilmarsson2021} & Effelsberg & 1360 & 59314 & $-0.71$ & 0.22 & \phantom{00}0.6\phantom{00} & \phantom{00}1.1\phantom{00} & \phantom{0}28.1\phantom{00} & \phantom{00}19 & 0.63 \\
\citet{Nimmo2022} & VLBI & 1400 & 59315 & $-1.06$ & 0.28 & \phantom{00}0.6\phantom{00} & \phantom{00}0.9\phantom{00} & \phantom{00}6.6\phantom{00} & \phantom{00}18 & 0.89 \\
\citet{Kirsten2024} & Onsala & 1400 & 59327 & $-0.63$ & 0.26 & \phantom{0}13.3\phantom{00} & \phantom{0}13.3\phantom{00} & 693\phantom{.000} & \phantom{000}8 & 1.00 \\
This work & uGMRT & \phantom{0}400 & 59366 & $-0.80$ & 0.40 & \phantom{0}40.0\phantom{00} & \phantom{0}40.0\phantom{00} & 679\phantom{.000} & \phantom{000}6.0 & 1.00 \\
\hline
\\[0.1cm]
\multicolumn{11}{c}{Fall 2021} \\
\hline
%\\[0.1cm]
% AB: previous version did not group pulses closer than 100 ms
\citet{Zhang2022} all & FAST & 1250 & 59483 & $-2.00$ & 0.20 & \phantom{00}0.010 & \phantom{00}1.69\phantom{0} & \phantom{0}10.4\phantom{00} & \phantom{0}696 & 0.15 \\
\citet{Zhang2022} & FAST & 1250 & 59483 & $-0.49$ & 0.09 & \phantom{00}0.020 & \phantom{00}0.020 & \phantom{00}5.4\phantom{00} & \phantom{00}31 & 1.00 \\
\citet{Zhang2022} & FAST & 1250 & 59484 & $-2.21$ & 0.83 & \phantom{00}0.012 & \phantom{00}1.19\phantom{0} & \phantom{00}3.5\phantom{00} & \phantom{00}63 & 0.14 \\
\citet{Zhang2022} & FAST & 1250 & 59485 & $-1.64$ & 0.27 & \phantom{00}0.010 & \phantom{00}1.61\phantom{0} & \phantom{0}10.4\phantom{00} & \phantom{0}190 & 0.20 \\
\citet{Zhang2022} & FAST & 1250 & 59486 & $-1.80$ & 0.18 & \phantom{00}0.013 & \phantom{00}1.16\phantom{0} & \phantom{00}5.6\phantom{00} & \phantom{0}412 & 0.26 \\
\hline
\\[0.1cm]
\multicolumn{11}{c}{Winter 2022} \\
\hline
\\[0.1cm]
\citet{Kirsten2024} & \mbox{Wb-RT1} & 1380 & 59603 & $-1.64$ & 1.16 & \phantom{0}13.4\phantom{00} & 224\phantom{.000} & 699\phantom{.000} & \phantom{00}24 & 0.17 \\
\citet{Kirsten2024} & Onsala & 1400 & 59612 & $-2.75$ & 2.75 & \phantom{0}14.9\phantom{00} & \phantom{0}67.7\phantom{00} & 122\phantom{.000} & \phantom{000}6 & 0.50 \\
This work & WSRT & 1369 & 59612 & $-0.71$ & 0.25 & \phantom{00}4.2\phantom{00} & \phantom{00}4.2\phantom{00} & 215\phantom{.000} & \phantom{00}10.0 & 1.00 \\
This work & uGMRT & \phantom{0}650 & 59617 & $-3.59$ & 1.47 & \phantom{0}11.0\phantom{00} & \phantom{0}26.0\phantom{00} & \phantom{0}59.0\phantom{00} & \phantom{00}10.0 & 0.80 \\
\citet{Kirsten2024} & Stockert & 1381 & 59619 & $-1.72$ & 0.65 & \phantom{0}11.0\phantom{00} & \phantom{0}36.1\phantom{00} & 250\phantom{.000} & \phantom{00}13 & 0.69 \\
\\[0.1cm]
\hline 
\label{table:PLI_all}
\end{tabular} 
\tablefoot{The columns are: the reference paper, telescope, central 
frequency  $\nu_\mathrm{c}$, MJD of first FRB detected, measured power-law index 
 $\am$ with its error $\alpha_\mathrm{err}$, minimum fluence $F_\mathrm{obs}$ 
 of bursts recorded, optimal minimum fluence $F_\mathrm{PL}$ for PL fitting, 
 maximum fluence $F_\mathrm{obs}$ recorded, total number of bursts, 
 and the fraction of bursts in a PL-like tail.}
\end{center}
\end{table*}

%%%%%%%% PLI comparison plot - Zhang only
\begin{figure}
\centering
   \includegraphics[width=0.4\textwidth]{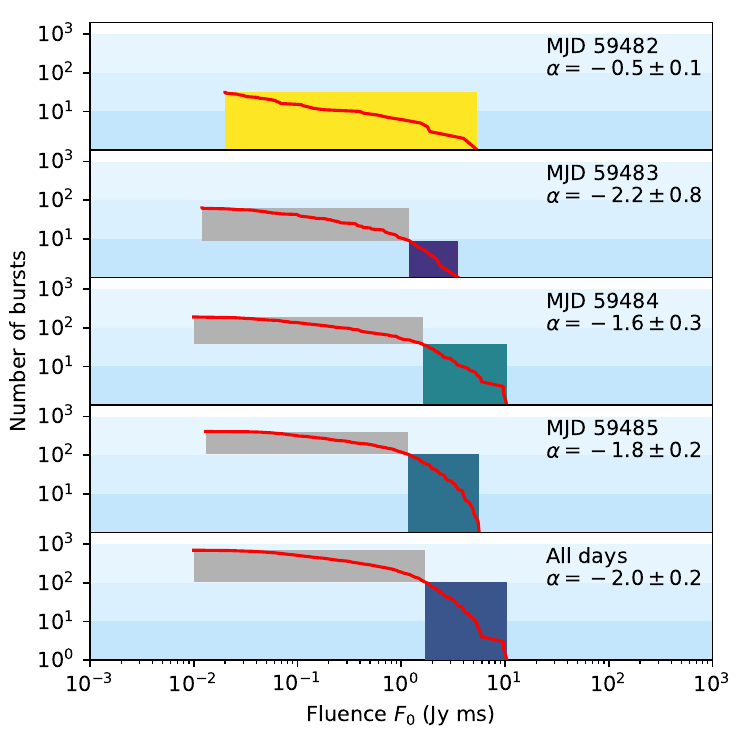}
    \caption{Un-normalized survival function of the fluence distributions (red line) 
    for the bursts recorded at the end of \Fall\ {activity window} \citep{Zhang2022}. The gray 
    rectangles mark the regions in the fluence--burst number parameter space that contain 
    bursts excluded from the PL fit (see Section~\ref{subsec:fluence_fit} for details). 
    The colored rectangles contain bursts included in the fit;  the color represents 
    the steepness of the best-fit {power-law index} (PLI) ranging from light yellow ($\am=-0.5$) to dark 
    blue ($\am=-2.5$). Light blue bands of progressively lighter shades in the 
    background correspond to regions containing 10, 100, and 1000 bursts. }
\label{fig:PLI_Zhang}
\end{figure}

%%%%%%%% PLI comparison plot all lit
\begin{figure}
\centering
\includegraphics[width=0.5\textwidth]{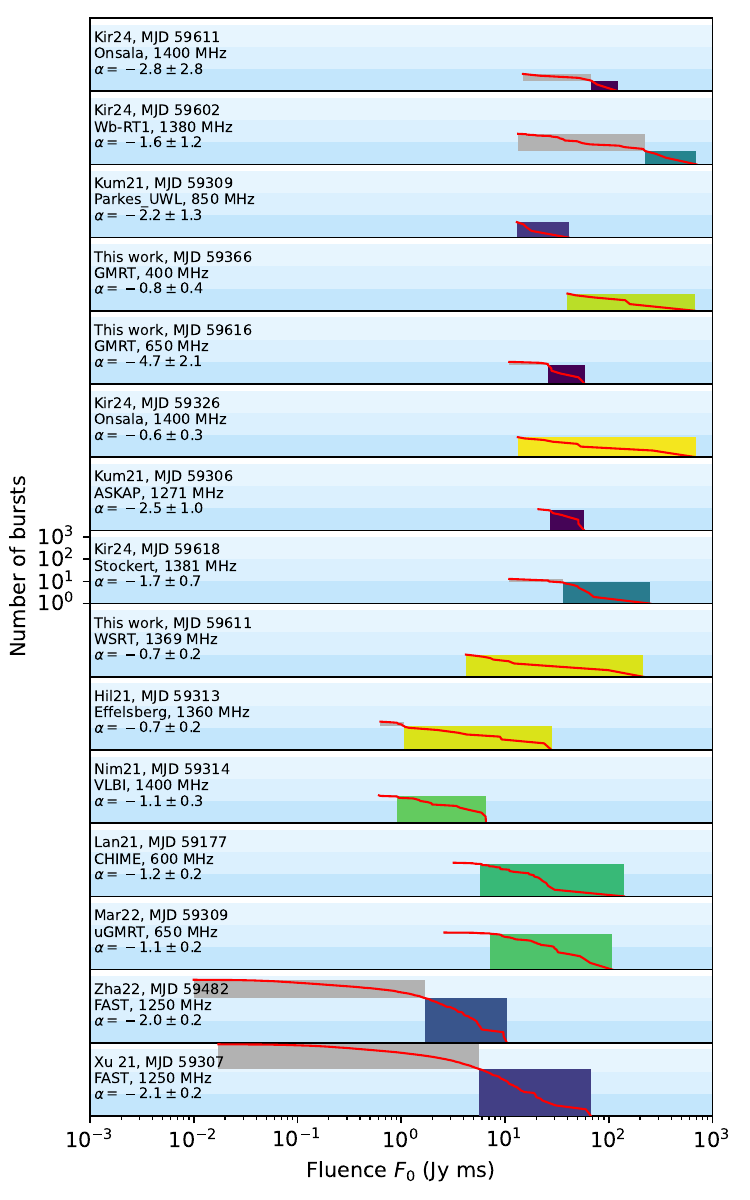}
    \caption{Un-normalized survival function of the fluence distributions for 
    the bursts recorded in the previous studies and current work (see caption 
    of Fig.~\ref{fig:PLI_Zhang} for the explanation of the legend and caption 
    of Fig.~\ref{fig:fluence_distr} for the literature abbreviations). 
    Here the burst samples are ordered by the number of bursts in 
    the best-fitting  PL tail. MJD refers to the start of respective observing campaigns.}
\label{fig:PLI_all}
\end{figure}

As an illustration, we refit the  fluence distributions  recently obtained by 
\citet{Kirsten2024} using both methods: graphical and MLE. The outcome of the 
original fits prompted the authors to speculate that there may be two 
separate populations of pulses emitted by \src, with the more energetic 
population having substantially flatter distribution. Unlike the authors, 
we fit fluences instead of spectral energy densities{, however
this does not affect the shape of the distribution}. While we use all 
available Onsala measurements, we necessarily exclude the Stockert observations 
from fitting with both methods, as the large difference in exposure time 
and sensitivity makes combining the  datasets for subsequent MLE fitting
very difficult. As in \citet{Kirsten2024}, the optimal minimum fluence value 
for both the FAST and Onsala distributions was determined with
\texttt{powerlaw}.\footnote{\url{https://pypi.org/project/powerlaw/}, 
modified according to Eq.~\ref{eq:a_unbiased}-\ref{eq:a_err_unbiased} in 
Appendix~\ref{app:PL_fit}{, see also \citet{James2019}}.} 
{This software finds the optimal minimum
fluence by minimizing the Kolmogorov-Smirnov distance between the data and the fit 
on a sample of fits that start from each unique value in the dataset.}
{In the same way as \citet{Kirsten2024}, the obtained minimum fluence value was 
used to filter fluences for fitting with the graphical method.}

Despite the above-mentioned differences in approach, our implementation of 
the graphical method reproduces the values of the measured {power-law index} (PLI) $\am$ and 
its corresponding bootstrapping errors as reported by \citeauthor{Kirsten2024}: 
$\am(\mathrm{Onsala}) = -0.50 \pm 0.06$ and $\am(\mathrm{FAST}) = -2.11 \pm 0.09$.\footnote{Bootstrapping errors are generally not well-suited for determining PLI uncertainties, see Appendix~\ref{app:PL_fit}.} 

{The MLE method, however, results in $\am(\mathrm{Onsala}) = -0.63 \pm 0.26$ and 
$\am(\mathrm{FAST}) = -2.15 \pm 0.19$. Clearly, the MLE determines the errors to the fit 
to be larger than the graphical method indicates.}
{In order to determine the statistical significance of the difference between 
the Onsala and FAST measurements we compared this difference to the respective 
measurement errors added in quadrature. This method assumes that both measured values are 
drawn from normal distributions, with a mean equal to the true value and a variance defined by 
measurement error. Therefore the difference between measurements is also normally 
distributed, with a mean of 0 and a variance equal to the sum of variances of individual 
measurements. In our case, for the graphical method the ratio of the measurement difference 
to its supposed standard deviation is 13.7, corresponding to practically zero chance 
probability. For MLE, the ratio is 4.7, corresponding to a chance probability of 
$1.3\times10^{-6}$. This indicates that the difference in PLI between Onsala and 
FAST is much less significant when measured with the MLE method. 
}

However, the distribution of $\am$ around true value $\at$ is not 
Gaussian: $\am|\at$ obeys a gamma distribution, which deviates from 
a normal distribution for small sample sizes, $\Ns\lesssim 100$. For 
a uniform prior on $\at$  this results in the posterior distribution for 
$\at|\am$ being skewed toward steeper PLI for small $\Ns$. In 
Fig.~\ref{fig:Kirsten_distr} we show the posterior distributions
for both the FAST and Onsala samples constructed from the fitted 
$\am$ on a grid of trial $\at$. The posterior $\at$ is well described 
by the gamma distribution from \citet{James2019} modified for the 
unbiased estimate $\alpha'$ (in their notation) by taking $M=\Ns-1$:
\begin{equation}
 p(\at|\am) \sim (-\at)^{\Ns-1}\left(\frac{\Ns-1}{-\am}\right)^{\Ns} \exp\left(-\dfrac{\am(\Ns-1)}{\at}\right).
 \label{eq:post_prob}
\end{equation}
Having two posterior distributions, one may calculate the probability 
that the difference between the  $\at$ values for the FAST and Onsala samples
is smaller than some threshold value (Fig.~\ref{fig:Kirsten_distr}). 
These probabilities remain low for differences in PLI less than 
$\sim 0.5$, indicating that a significant difference exists 
between $\at(\mathrm{FAST})$ and $\at(\mathrm{Onsala})$. This significance is, 
however, appreciably smaller than implied by graphical method. For example,
 $p(\Delta\at\leq0.5) = 5.3\times10^{-4}$  graphically, but $3.2\times10^{-2}$ 
according to the  MLE.

To summarize our findings up to here, accurate estimates of the PLI for 
small burst samples critically depend on two factors: first, using an 
unbiased PLI fitting method, and second, taking into account the skewness of 
the posterior probability distribution.

There is, however, one more caveat connected to sampling from flattened distributions.
If the observed fluence distribution  flattens at the low  end, either due 
to instrument sensitivity, or intrinsically, then this flattening cannot 
always be recognized in small samples, and the measured power-law index (PLI) 
can be biased toward shallower values. The amount of this bias depends on 
the extent of flattening. As an example, we show the distribution of 10-pulse PLIs
from the FAST \Spring\ sample in Fig.~\ref{fig:flattening}. The sample was 
truncated at 5\,Jy\,ms and the flattening signifies an intrinsic property
of the pulses. The PLI distribution is skewed toward shallower values, 
which is expected because low-fluence pulses with shallower survival 
functions are more abundant. However, about 17\% of these 10-pulse samples 
do not exhibit apparent flattening at the lower-fluence edge of survival function.
Their survival functions are well-fitted with a single PL for the whole
extent of the 10-sample distribution. A chance observation resulting in one 
 such 10-burst realization may well lead to the erroneous conclusions 
 that the distribution of burst fluences obeys a PL with a shallow index,
and to a subsequent unfounded scientific interpretation.

The exact amount of PLI bias toward shallower values depends on the sample 
size and on the underlying fluence distribution as observed with a specific 
instrument under a specific setup. Assumptions about intrinsic
fluence distribution as well as  models of instrumental limitations
\citep[see, e.g.,][for such models]{Gardenier2019,Wang2024}
should both be taken into account while interpreting the measured PLIs. 

\subsubsection{Fitting results}

In this section, we review the PLI measurements for the fluence survival functions 
of bursts recorded in our observations and reported in the literature. 
Table~\ref{table:PLI_all} summarizes the results of the MLE fitting procedure. Unlike 
\citet{Kirsten2024}, we did not combine bursts recorded by different telescopes, and 
we did not correct for the  unknown possible underestimation of CHIME 
fluences \citep{Lanman2022}. For the FAST sample and our WSRT observations, 
we combined bursts with separation smaller than 100\,ms to facilitate comparison 
with \citep{Kirsten2024}. Such proximity threshold lies within the 
short-recurrence component of the bimodal waiting time distribution and is 
smaller than the 400 ms sub-burst separation threshold in \citet{Zhou2022},
which was derived from the trough between the two components of the waiting time distribution. 
In practice this means that for the small fraction of bursts the fluences 
of individual sub-burst clusters, as defined by \citep{Zhou2022} were counted separately.
This, however, had only minor influence on the shape of fluence distributions.

In Table~\ref{table:PLI_all}, all measured $\am$ values are within the 
range of $-0.5$ to $-4$, but with the caveat that nominal PLI values,
even with their respective MLE fit uncertainties, should be treated with caution 
due to the apparent flattening of distributions at the lower-fluence end, which
can be either instrumental or intrinsic. To illustrate this, we show the 
\texttt{powerlaw} fit for four individual sessions of the \Fall\ FAST burst sample 
(Fig.~\ref{fig:PLI_Zhang}). During the observing run, the rate of pulses increased 
more than tenfold, from 31 to 412 combined bursts per one-hour session. The overall 
shape of the fluence distribution did not change dramatically, yet for the first 
session, the PL fit yielded a flat $\am=-0.49\pm0.09$ without a low-fluence cutoff. 
For the subsequent sessions, the optimal fit excluded the low-fluence region, 
resulting in a much steeper PLI {(e.g., $\am=-1.8 \pm 0.2$ for MJD 59485)}. It is possible that the shallow $\am$ for the first 
session is solely due to the aforementioned bias present in small samples drawn 
from a flattened distribution; however, intrinsic variability cannot be ruled out.

The fits for per-session fluence distributions of the FAST sample from the \Spring\ 
{activity window} also exhibited a dependence on the number of bursts recorded per session. 
The number of bursts varied from 11 to 99 per session. The measured PLIs were more 
diverse for smaller burst samples, ranging from approximately $-4 \pm 2 $ to $-0.7 \pm 0.3$.\footnote{{These uncertainties are indicative, the actual values vary by about 50\% for similar $\am$ measurements on different observing days.}} 
{Occasionaly, $\am$ reached much steeper values of about $-10 \pm 9$.} The fraction of power-law-obeying bursts 
ranged from 5\% to 90\%, with steeper indices corresponding to larger minimum 
fluences and a smaller number of bursts in the PL tail. The two most 
prolific sessions, with more than 80 FRBs each, yielded $\am$ values of $-1.2 \pm 0.3$ and 
$-1.7 \pm 0.3$, with 25\% of bursts belonging to the PL tail.

All three burst subsamples in our study are small, with ten or fewer bursts each. 
The bursts from two GMRT sessions in the \Spring\ and \Winter\ {activity window}s 
resulted in dramatically different $\am$ values, $-0.80 \pm 0.40$ and $-3.59 \pm 1.47$. The 
steep PLI resembles similar steep values of the Spring FAST per-session 
fit when the burst  sample size was small,  and a relatively large minimum
fluence was found for the PL tail. 
The WSRT burst sample is described by a shallow distribution with $\am = -0.71 \pm 0.25$ 
and no apparent low-fluence flattening. This PLI is close to the ones reported 
by \citet{Kirsten2024} and \citet{Hilmarsson2021}, but extends to smaller fluences.
Our $\am$ for the CHIME data is shallower than 
the MLE fit from \citet{Lanman2022}, as they 
excluded the brightest burst due to uncertainties in fluence determination. 

Fig.~\ref{fig:PLI_all} provides a graphical representation of the PLI 
measurements from Table~\ref{table:PLI_all}. Burst samples are sorted by the 
number of bursts in the PL tail. Overall, PLIs determined from burst samples 
smaller than about a hundred tend to have diverse values that are inconsistent 
between different studies. For the largest sets of pulses (696 and 1715 FRBs 
from the two FAST studies), only a small fraction (about 10\%) of the brightest 
pulses follow a power-law distribution, with $\am$ close to $-2$ and 
minimum fluence differing by a factor of 3. Since these 
observations were performed under the same observing setup, the difference 
signifies intrinsic variability between {activity window}s.

Given the inevitable instrumental bias, possible intrinsic frequency evolution 
and temporal variability, constraining the PLI is a daunting task. One can also well ask 
the question whether PL approximation should be used in this case at all,
since for both of the two samples of bursts that were largest and most sensitive,  
only 10\% of bursts have fluences that obey a PL distribution.

\subsubsection{Fitting code availability}

To facilitate the future use in the community of the correct, MLE algorithms for fitting burst fluences,
we have made available an ipython notebook that implements the various types of PL fits.
It includes instructions on how to adapt the \texttt{powerlaw} package 
as described in Appendix~\ref{app:PL_fit}.
The notebook is hosted on 
Zenodo\footnote{\scriptsize\url{https://doi.org/10.5281/zenodo.12644702}}
and GitHub.\footnote{\scriptsize\url{https://github.com/TRASAL/FRB_powerlaw}}

% Also see acks

%%%%%%%%%%% Scintillation
\begin{figure}
    \centering
    \includegraphics[width=0.4\textwidth]{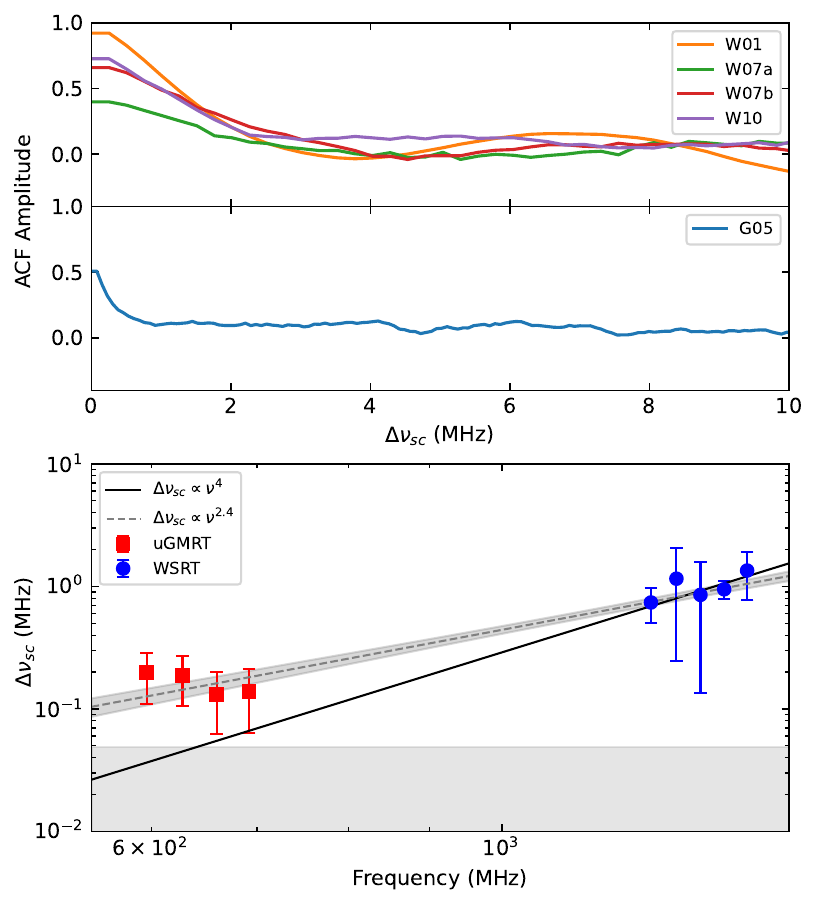}
    \caption{Decorrelation bandwidth analysis. The top panel shows 
    the ACF from 1220\,MHz to 1453\,MHz of the Apertif bursts W01 
    (orange), W07a (green), W07b (red), W10 (purple), and G05 (blue). 
    The middle panel shows the ACF from 580 to 709\,MHz of the uGMRT 
    burst G05 in blue. The bottom panel shows how the decorrelation 
    bandwidth evolves with frequency. The Apertif bandwidths 
    were divided into five sub-bands, and the uGMRT bandwidth into four sub-bands. 
    The gray dashed line with a shaded area shows the best-fitting 
    power law, with an index $\alpha=2.4\pm0.2$. As a reference, 
    the black solid line shows an index $\alpha=4$ expected from 
    a thin screen. The gray shaded region indicates decorrelation 
    bandwidths that cannot be resolved by the uGMRT resolution.}
    \label{fig:scintillation}
\end{figure}

\subsection{Scintillation}

%%%%%%%% Fig: TOA statistics
\begin{figure}
    \includegraphics[width=0.5\textwidth]{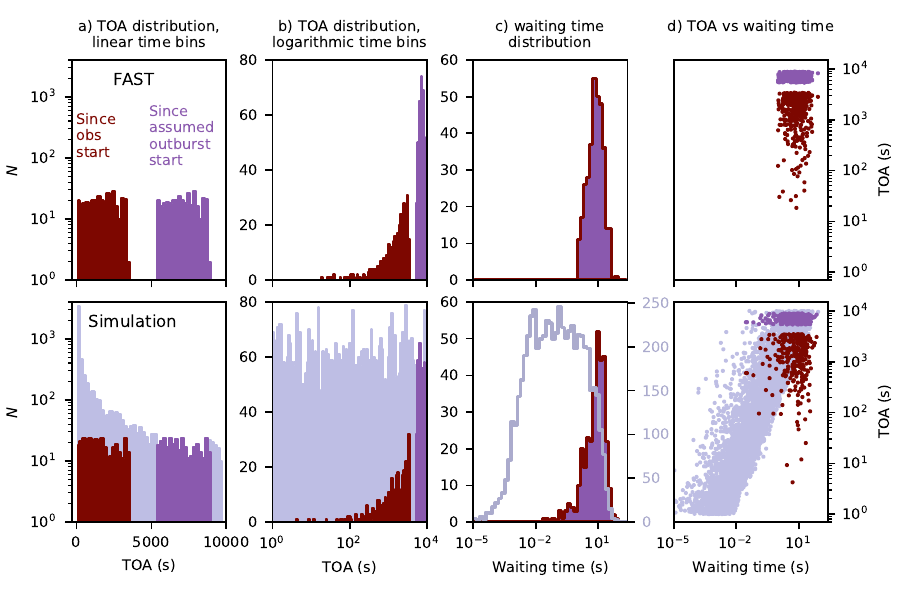}
    \caption{TOA and waiting time distribution for the most prolific 
     \src\ observing session from \citet{Zhang2022} on MJD 59486 (\textit{upper row}), 
    and a series of log-uniform random variables (\textit{lower row}) that 
    simulate an FRB outburst. In our model only a small fraction of the 
    outburst duration is covered by observations,
    long after the burst onset. The columns are: \textit{a)} distribution of 
    TOAs measured from the first burst arrival (brown) and from the 
    hypothetical  outburst start  (violet), with the pale shade of violet 
    showing the whole underlying mostly unobserved outburst; \textit{b)} 
    same distribution, but binned 
    with logarithmic time bins; \textit{c)} distribution of TOAs vs 
    waiting time between two bursts; \textit{d)} distribution of waiting 
    times. Bursts separated by less than 400\,ms
    combined, producing a single TOA for the start of the combined burst.}
    \label{fig:waiting_time}
 \end{figure}

In order to measure the decorrelation bandwidth, we obtained the spectra of
the brightest bursts in our sample, G05, W01, W07A, W07b, and W10, by averaging 
their emission in time over the time
bins where the signal is four times larger than the noise standard deviation.
Note that for W07a, only the last, brightest component satisfies this criterion. 
For each ALERT burst, we removed the data above $\sim1423$\,MHz, where RFI 
becomes strong in the observations. For the uGMRT burst G05, we fitted the
burst spectrum to a Gaussian and only took the frequencies within two 
standard deviations of the center in order to have enough signal for the 
analysis; this is between 580\,MHz and 709\,MHz. 
Next we compute the ACFs 
of the spectra, removing the zero-lag frequency value, and fit the central 
peak of the ACF to a Lorentzian. The decorrelation bandwidth is often 
defined as the half-width at half-maximum of the ACF's fitted Lorentzian 
\citep[see, e.g.,][Section 4.2.2]{Lorimer2005}. For the ALERT bursts, we 
obtain an average decorrelation bandwidth (weighted by the inverse standard 
deviation of each measurement) \scbw\,$=1.1\pm0.2$\,MHz at the central 
frequency 1336.5\,MHz. For G05, we obtain \scbw$=0.15\pm0.05$\,MHz at
the central burst frequency, 644.5\,MHz. {Both \scbw\ values
are much larger than the channel width of 195 and 48\,kHz, respectively}

The frequency-dependent intensity variations produced by scintillation 
are expected to follow a power law evolution of the form \scbw\,$=A\nu^{\gamma}$, 
with $\nu$ the frequency in GHz, $A$ a constant that gives the decorrelation 
bandwidth in MHz at 1\,GHz, and $\gamma$ the scintillation index, expected 
to be $\gamma=4$ for scintillation produced by a thin screen and $\gamma=4.4$ 
for scintillation produced in an extended, turbulent medium. To measure the power-
law index of the decorrelation bandwidth, we divide the bandwidth into 
several sub-bands, and we measure the half width at half maximum (HWHM) 
in each sub-band as described above. 
We divide the Apertif bursts into five sub-bands, and the uGMRT one into four.
For the Apertif bursts, we compute the HWHM weighted average in each 
frequency sub-band, and then we fit all decorrelation bandwidths as a function 
of frequency to a power-law spectrum. We obtain $A=0.44\pm0.03$\,MHz, and $\gamma=2.4\pm0.2$.
The results from the scintillation analysis are presented in Fig.~\ref{fig:scintillation}.

Our measurements at both frequencies are based on a small number of bursts. 
Previous studies have shown that both the decorrelation bandwidth and $\gamma$ 
vary substantially when measured on individual bursts within a relatively 
narrow bandwidth of 500\,MHz centered around 1250\,MHz: \citet{Xu2021} 
reported a mean $\gamma$ of 4.9 with a standard deviation of 6.4. This 
variation may be at least partially influenced by the intrinsic spectral 
structure of the bursts. Inferring $\gamma$ from the combined spectra of 
a few dozen bursts detected within an hour-long observing session resulted 
in a  shallower $\gamma$ of $3.0\pm0.2$ \citep{Zhou2022}.

\citet{Main2022} increased the robustness of $\gamma$ determination by 
comparing decorrelation bandwidths at 700 and 1400\,MHz. Their setup is 
similar to ours in terms of frequency coverage, but their sample of bursts 
is larger. They reported a power-law index of $\gamma=3.5\pm0.1$, which 
is also shallower than the one for the thin screen model. However,  
both in our measurements and those from \citet{Main2022}, the decorrelation 
bandwidth at the lowest radio frequencies may be biased by insufficient 
frequency resolution, resulting in $\gamma$ being biased toward  
shallower values.

The study of \citet{Main2022} utilized bursts from the \Spring\ 
{activity window}, with low- and high-frequency observations conducted several days 
apart. Our measurements at 1.4\,GHz come from the \Winter\ {activity window}, 245 days 
after the \Spring\ observations at 650\,MHz. \citet{Wu2024} reported small 
annual variations in the observed decorrelation bandwidth attributed 
to the Earth's movement with respect to a moderately anisotropic scattering 
screen located in the Milky Way \citep[see also][]{Main2022}. This 
variation is much smaller than the scatter associated with measurements 
from individual sessions and therefore cannot be the main source of 
systematic uncertainty in our $\gamma$ measurement. The available 
bulk of observational data does not show evidence for a secular 
trend in the decorrelation bandwidth or $\gamma$ across three  
{activity window}s.

\cite{Main2022} detected enough nearby bursts from \src\ to measure its 
scintillation timescale of $13.3\pm0.8$\,min in the L-band. Unfortunately, 
the most closely spaced bursts with sufficient S/N in our sample 
(W01 and W07) are separated by $\sim110$\,min, which does not allow us 
to probe the scintillation timescale of the source. The correlation 
coefficient between W01 and W07a/b is close to zero, consistent with 
the previous measurement.

\subsection{TOA statistics} 
\label{subsect:TOA}

The similarity in burst arrival statistics between FRBs and the high-energy 
short bursts from magnetars was one of the key observational facts put 
forward by
\defcitealias{Wadiasingh2019}{WT19}\citet[][hereafter: \citetalias{Wadiasingh2019}]{Wadiasingh2019}
in support of their  crust motion and low-twist model of FRB generation.
Both observed burst types feature a  log-normal distribution
of waiting times, and a log-uniform distribution of TOAs 
as measured from the outburst start time. \citetalias{Wadiasingh2019} 
studied the distribution of burst arrival times using a sample of 
93 FRBs from \srcRi\ recorded over the span of five hours. Some of the FAST 
observations of \src\ provide much higher a pulse rate (up to a few 
hundred pulses per hour), but the duration of the observing session is smaller,
typically less than two hours. 

Using the publicly available information about burst fluences and 
arrival times from \citet{Xu2021} and \citet{Zhang2022}, we constructed 
the distribution of waiting times between successive bursts. As mentioned before,
this distribution  consists of two non-overlapping log-normal 
components. Based on a common definition of sub-bursts, we combined  
bursts which were separated by less than 400\,ms into one, using an iterative routine. 
For each resulting cluster of sub-bursts, the TOA was taken to be the time of arrival of 
the first burst, and the cluster fluence to be the sum of the individual component 
fluences. Combined bursts from each observing day were analyzed separately. 

Figure~\ref{fig:waiting_time} shows the distribution of burst TOAs measured
with respect to the session start for the most prolific observing session, 
on September 29 2021 (MJD 59486; see Table~\ref{table:PLI_all}).
The distribution is uniform with linear time bins and, 
unlike for \srcRi\ \citepalias{Wadiasingh2019},  there is no linear
correlation between the logarithms of the TOAs and the corresponding waiting times
(the brown points in the top-right subplot). This discrepancy can, however, be 
explained if the observation took place some time 
after the outburst started, when the burst rate no longer changed significantly 
during a session.

To illustrate this, we simulated burst arrival times by generating $6\times10^4$ 
log-uniformly distributed random variables $t$, using \texttt{scipy.stats.loguniform}. 
These random variables fell within the range of $1<t<10^4$\,s with a probability 
density function of $p(t) \sim 1/t.$ Approximately 1.5 hours after the start 
of the simulated outburst, the rate of occurrence for $t$ remains relatively 
constant over the duration of the FAST session.

If TOAs are measured from the start of the simulated outburst, $\mathrm{TOA} = t-5400$\,s, 
their statistics closely resemble the statistics of real bursts 
(Fig.~\ref{fig:waiting_time}). This suggests the possibility that the observed 
bursts are part of an outburst that began several hours before the observations. 
In this model, the variation in the session-to-session burst rate, which shows 
significant growth toward the end of the \Fall\ {activity window}, could 
be attributed to different outbursts overlapping, or occurring in close 
succession. Unfortunately, the scope of this work does not permit a quantitative 
assessment of the probability of FAST missing the start of an outburst after 
observing \src\ for almost 60 days with $<10\%$ duty cycle. Such an assessment 
would require an estimation of the tentative duration of an outburst and the 
frequency of their occurrence.

%%%%%%%% Frequency-dependent {activity window}
\begin{figure*}
\centering
\includegraphics[width=1.0\textwidth]{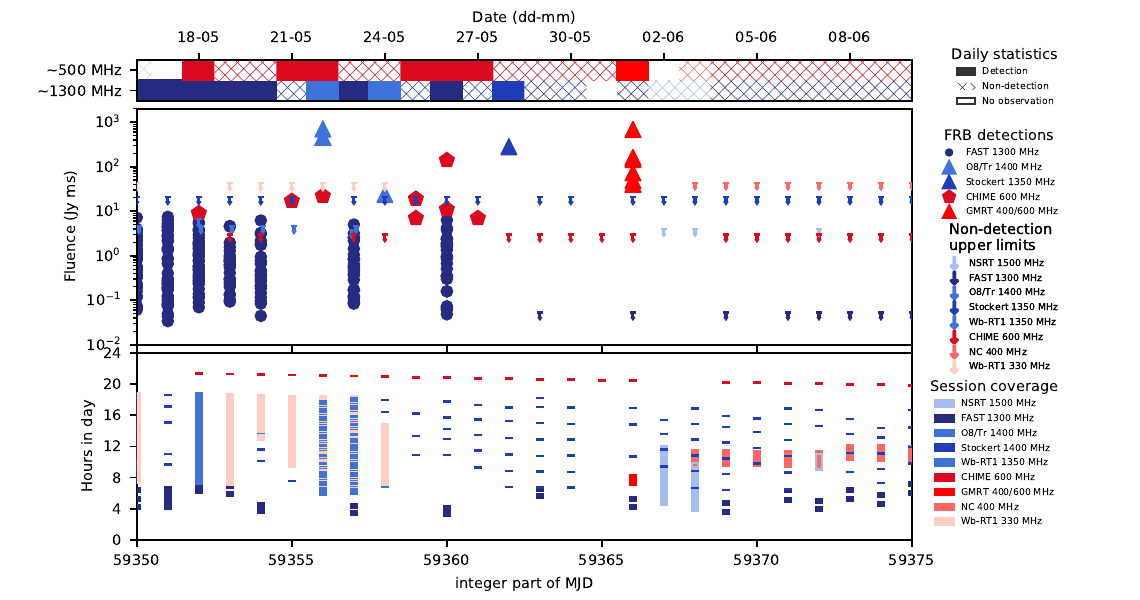}
\caption{\textit{Upper panel:} Overview of daily detection statistics 
at the end of \Spring {activity window}
from all telescopes combined. A filled rectangle indicates that the 
source was detected, a hatched rectangle for non-detected, and a white rectangle signifies there were
no observations that day. Lower frequencies are shown at the top, following 
Sect.~\ref{sec:geo:rfm}. \textit{Middle panel:} FRB fluences, 
together with upper limits for non-detections vs   the integer part of the 
observation MJD. The time region around the end of the \Spring\ activity 
window is shown. The red triangles mark  GMRT observations at 400 or 
650\,MHz (bursts G01--G06 from Fig.~\ref{fig:FRBs_GMRT}). The darker and 
lighter red shades show detections (pentagons) or upper limits for 
bursts in similar frequency regions from the observations by 
CHIME/FRB \citep{Lanman2022} and NC \citep{Trudu2022}. The results 
of higher-frequency L-band observations by FAST \citep{Xu2021} and 
NSRT \citep{Mao2022} are shown in darker and lighter blue, 
with dots marking FAST detections. 
Detections by Torun/Onsala/Stockert (T/O/S) dishes from \citet{Kirsten2024} 
are shown as blue triangles (see text for details on fluence 
upper limits). \textit{Lower panel:} Observation time coverage.}
\label{fig:exposure}
\end{figure*}

\subsection{Frequency-dependent {activity window}}
\label{sec:fdac}

\src\ was extensively observed, by multiple telescopes, around the 
end of the \Spring\ {activity window}. Figure~\ref{fig:exposure} provides 
a summary of the time/frequency coverage of these observations, including 
the TOAs and fluences of the  detected FRBs, plus fluence upper limits 
for any non-detections.

The majority of the FRB detections in the L-band were provided by sensitive 
FAST observations. These observations occurred at intervals of 1 or 
3 days, with each 2-hour session yielding the detection of dozens of 
bursts with fluences above 0.053\,Jy\,ms. Notably, there was an abrupt 
cessation of emission between MJD 59360 and 59363, after which no 
FRBs were recorded despite the unchanged observing setup and cadence.

These FAST observations were complemented by \citet{Kirsten2024}, who 
conducted near-daily observations using several smaller telescopes. 
Their observations necessarily featured a significantly higher detection threshold, 
approximately 10\,Jy\,ms. The last known pulse from the  \Spring\ session, 
detected by \citeauthor{Kirsten2024}, occurred on May 28, one day before the 
first non-detection by FAST, on a day when FAST was not conducting observations.

Shortly after this then yet unknown quenching, \citet{Mao2022} executed an 
extensive targeted search for bursts from \src\ in the L-band using the 
Nanshan 26 m radio telescope (NSRT). The authors determined a minimum detectable 
fluence of 4\,Jy\,ms. On June 02 and 03, when FAST was not observing, 
the NSRT sessions were significantly longer than those generally used at FAST.
If the source had persisted as active as before, several bright pulses 
should likely have been detected on these dates -- but none were. On June\,07,
both NSRT and FAST observed {\src}, a few hours apart, with neither telescope 
detecting any bursts.

At the lower frequencies centered around  400 to 600\,MHz, the majority of 
observations are provided by CHIME/FRB. That transit instrument records 
\src\ for 3.13 minutes virtually every day in the frequency range of 
400--800 MHz. \citet{Lanman2022} estimate the burst rate after March 20 to 
be between 0.9--2 $\times$ 100 day$^{-1}$, and the bursts exhibit Poissonian repetition. 
No bursts were detected in the five sessions  between May 27 and our GMRT 
detections described below; this absence has a Poissonian probability of between 0.11 
and 0.37, assuming constant observing and instrument conditions.
Beyond purely Poissonian variations, it is noteworthy that the rate may also 
intrinsically vary on timescales shorter than a month, as indicated by the apparent 
inconsistency between the rate derived from CHIME/FRB observations and that from a 
3-hour April session at 550--750 MHz at GMRT \citep{Marthi2022}. For the CHIME/FRB 
fluence limit, we take the lowest-fluence burst detection from \citet{Lanman2022}, 
namely 3\,Jy\,ms.

\citet{Trudu2022} observed \src\ at similar frequencies, 400--416\,MHz,
with the Northern Cross (NC) radio telescope, for 68 hours in April and June 2021. 
The authors estimate a minimum detectable fluence of 44\,Jy\,ms and expect
$1\pm1$ bursts to be detectable over this whole observing campaign.
That expectation is based on  the rates and the power-law fluence
distribution slope from CHIME/FRB monitoring \citep{Lanman2022}, and assuming there is no 
burst rate variability. Similarly, \citet{Kirsten2024} observed 
\src\ at 350\,MHz with Wb-RT1  during several days, right before the L-band quenching, 
with an estimated minimum detectable fluence of 42\,Jy\,ms, and not detecting any bursts.

Our observations with GMRT yielded six bursts on 2021 June 01, all of them 
with fluences above 40\,Jy\,ms and one burst reaching fluence of 680\,Jy\,ms. 
FAST observed \src\ a few hours before these GMRT observations and placed a stringent 
$0.05$\,Jy\,ms upper limit on the FRB fluences in L-band.

Taken together, these detections and strict upper limits show that
after it stopped emitting at L-band, \src\ continued to produce bursts at 400 to 600\,MHz.
The evidence is shown in Fig.~\ref{fig:exposure}. We estimate that the
low-frequency radio emission may have lasted for 3--6 days after bursts stopped in L-band.
The lower limit of 3 days comes from a scenario where the L-band quenching happened  
right before FAST observations on May 29, and lower frequency emission 
ended right after the GMRT observations on June 01. The upper limit of 6 days 
follows from assuming the high frequencies cease right after the last detected 
Stockert burst on May 28 and low-frequency emission persist up to the 
non-detection session at the Northern Cross telescope on June 03.

So far, only \srcRiii\ is known to  display a similar frequency-dependent activity 
window in a repeating FRB \citep{Pastor-Marazuela2021, Pleunis2021b}. This 
source has a well-determined activity period of 16.3 days \citep{CHIME2020}. 
From comparing the activity phase-resolved burst detection rate in simultaneous 
observations using WSRT/Apertif and LOFAR, and in earlier CHIME/FRB observations, 
the authors conclude that higher frequencies appear to arrive earlier in phase.
This trend was confirmed to extend to frequencies as high as 5\,GHz.
Between 150\,MHz and 5\,GHz the center of activity window shifts by about 6\,days
and the width of the window shrinks from 3.6 to 1.0 days  \citep{Bethapudi2023}.

For \src, the activity window is challenging to determine precisely because 
of the uneven observational coverage, but evidence suggests that its duration 
is highly variable, with the duration of the inactive sessions ranging from a 
few months to at least two years \citep{Lanman2022}, and the active stages 
spanning months. Targeted searches have ruled out a periodicity of up to 10 days 
\citep{Xu2021, Niu2022, Du2023}, but longer periods are not yet disproved.

\section{Geometric constraints on emission regions}
\label{sec:geo}

In the low-twist magnetar model of FRB generation, bursts are produced in the 
magnetosphere of a neutron star via a pulsar-like emission mechanism 
(\citetalias{Wadiasingh2019}). The exact nature of that invoked
mechanism remains a 
long-standing mystery despite a plethora of observational pulsar facts and decades 
of ongoing modeling efforts. Nonetheless, the key observational properties of 
pulsar radio emission can be explained by a pair of phenomenological models, 
known as the radius-to-frequency mapping \citep[RFM;][]{Cordes1978}
and the rotating vector model 
\citep[RVM;][]{Radhakrishnan1969}. These models produce constraints on the overall 
magnetospheric geometry and the location of emission regions, based on a set of 
plausible assumptions. Below we  apply RFM/RVM techniques to bursts from \src.

\subsection{Radius-to-frequency mapping (RFM)}
\label{sec:geo:rfm}
The RFM model postulates that radio waves decouple from magnetic field lines at a certain 
altitude, and propagate tangentially to the local magnetic field line at the 
decoupling point. We  call this the emission point, although radio waves 
may actually be produced elsewhere \citep{Philippov2020}. Lower radio 
frequencies are related to emission points situated higher in the magnetosphere, 
where the dipolar field has diverged further, offering a natural explanation for 
the broadening of the on-pulse window at lower frequencies that is observed 
in radio pulsars. In pulsars, the on-pulse activity window is thus frequency dependent,
and wider at lower frequencies. 

Now, the presence of similar frequency-dependent activity window behavior 
is evident in both \src\ and \srcRiii. Nevertheless, a clear distinction 
arises in terms of timescale: in pulsars, the widening amounts to fractions 
of a second, whereas for FRB sources, the frequency-dependent edge of 
the activity window extends over the course of days. However, the difference 
in spin phase remains consistent between radio pulsars and FRB repeaters if 
the latter are ULPs with spin periods ($P$) of weeks or longer.

Qualitatively, the frequency-dependent edge of the {activity window} can be 
explained as follows: while the rotation of the magnetar slowly moves
our LOS through the magnetosphere, 
we detect radio waves from field lines with footpoints in the active 
regions on the stellar surface. Plasma propagates along the active field line and
generates radio emission of continuously decreasing frequency as it moves 
outward. Since this emission is highly beamed and its frequency 
is altitude-dependent, at any given moment of time the observed broadband 
spectrum is composed of radio waves coming from a range of altitudes and 
originating from different field lines. Consider, for clarity, only two radio 
frequencies:  high and low. We, the observer, detect emission along our single LOS. 
But the two different frequencies originate from field lines
with footpoints that form two separate, different paths on the surface 
of the star. These two paths may cross the edge of the active region at different 
times. Fortuitously, the low-frequency path may stay longer in the active region, 
resulting in detection of low-frequency FRBs after the cessation 
of higher-frequency emission. Since the effect is purely geometrical and 
there are no constraints on the shape of the active region for \src, 
many exact lags between high- and low-frequency quenching are possible, 
including negative lags,  where low-frequency ceases first. The lags may also 
vary from one episode of activity to another. Once accumulated, 
the statistical distribution of lag magnitudes and signs may provide 
clues for the magnetospheric geometry configuration and the spread of the active 
regions on the stellar surface.

That is the qualitative description to introduce the underlying concepts.
In what follows we   constrain the location of observable 
field line footpoints quantitatively. We assume the external magnetic field to be dipolar, 
designating the angle between spin and magnetic axis as $\alpha$. Since no external 
constraints are known for $\alpha$ we take it to belong to a grid of trial 
values ranging from $5\degree$ to $90\degree$ and spaced by $5\degree$.

In a spherical coordinate system aligned with the spin axis, the observer LOS 
is defined by a pair of longitude/latitude angles. The latitude $\theta_\mathrm{ob}$ 
(also called viewing angle) does not change with time while 
the longitude corresponds to the spin phase at the moment of time 
$t$: $\phi_\mathrm{ob} \equiv 2\pi t/P$. In the absence of 
any external constraints, we review a grid of $\theta_\mathrm{ob}$ 
ranging from $5\degree$ to $90\degree$ with the spacing of $5\degree$. 

%%%%%%%% Footpoint loops
\begin{figure}
   \includegraphics[width=0.45\textwidth]{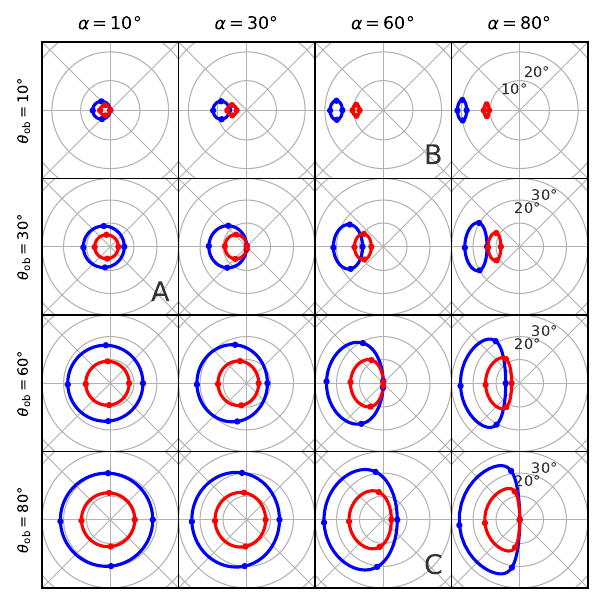}
   \caption{Maps of footpoint loops spanning all geometries.
Shown are the coordinates of the footpoints of the magnetic field lines which are 
potentially visible for a LOS characterized by viewing angle $\theta_\mathrm{ob}$, 
as it passes through a dipole field inclined to spin axis by an angle $\alpha$. 
The line colors reflect the observing frequencies shown in Fig.~\ref{fig:exposure}. 
The blue loop corresponds to emission seen at WSRT frequencies $\nu_\mathrm{hi}=1300$\,MHz,
coming from chosen emission altitude $r_\mathrm{hi}=5R_\mathrm{NS}$;
while  the red loop contains the footpoints visible in the lower GMRT 
band $\nu_\mathrm{low}=430$\,MHz, from $r_\mathrm{low}=15R_\mathrm{NS}$. The dots mark 
the quarters of the magnetar rotation. The coordinate system is centered on the magnetic 
dipole axis. The letters indicate the sets of geometry angles with emission regions plotted in 
Fig.~\ref{fig:Emission_regions_FL}.}
\label{fig:LOS_on_surface}   
\end{figure}

%%%%%%%% Emission region with field lines
\begin{figure}
\includegraphics[width=0.23\textwidth]{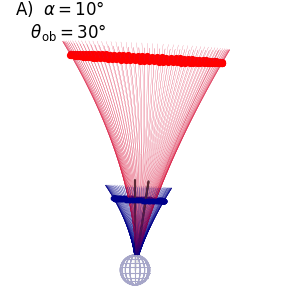}\includegraphics[width=0.23\textwidth]{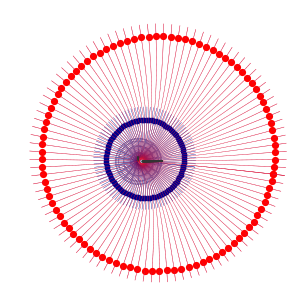}
\includegraphics[width=0.23\textwidth]{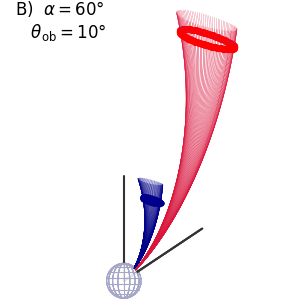}\includegraphics[width=0.23\textwidth]{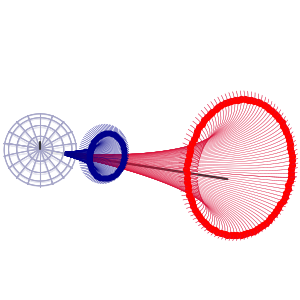}
\includegraphics[width=0.23\textwidth]{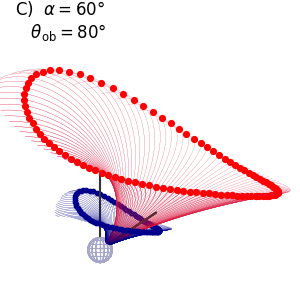}\includegraphics[width=0.23\textwidth]{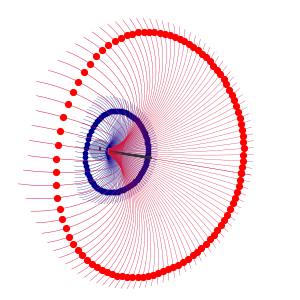}
\caption{Field lines that are potentially visible, for a number of 
neutron-star geometries. The geometry determines the LOS, which is 
characterized by the viewing angle $\theta_\mathrm{ob}$, and passes 
through a dipole field that is  inclined to the spin axis by an 
angle $\alpha$. These spin and magnetic axes are shown with thick black 
lines, and the NS coordinate lines are drawn in its rest frame.
The frequency-color mapping follows the previous figures. Field lines 
with radio emission observable at 430\,MHz are plotted in red,
those observable at 1300\,MHz in blue.  Dots mark the emission 
points, at an altitude of 15$R_\mathrm{NS}$ for  430\,MHz and at 
5$R_\mathrm{NS}$ for 1300\,MHz. The corresponding footpoint 
loops on the stellar surface are shown in Fig.~\ref{fig:LOS_on_surface}. 
\textit{Left column:} Side view with spin axis pointing vertically. \textit{Right column:}
Top view with spin axis pointing at the observer. \textit{Upper row:} Almost aligned
rotator, with our LOS passing close to the spin--magnetic axes. The emission points and field line 
footpoints form two almost concentric circles. \textit{Middle row:} Inclined 
rotator with a LOS passing away from the magnetic axis, but close to the spin axis. The LOS 
samples a restricted range of longitudes and latitudes on the stellar surface, and the emission 
points and footpoints loops are completely separate for the two frequencies, meaning 
that the lower-frequency loop is not encircled by the higher-frequency one. 
\textit{Lower row:} Inclined rotator with LOS passing away from the spin axis, 
but close to magnetic axis. The emission points and footpoints form concentric loops that 
are flattened on the side closer to the magnetic axis.}
\label{fig:Emission_regions_FL}   
\end{figure}

%%%%%%%% Active region cartoon
\begin{figure*}
\includegraphics[width=0.25\textwidth]{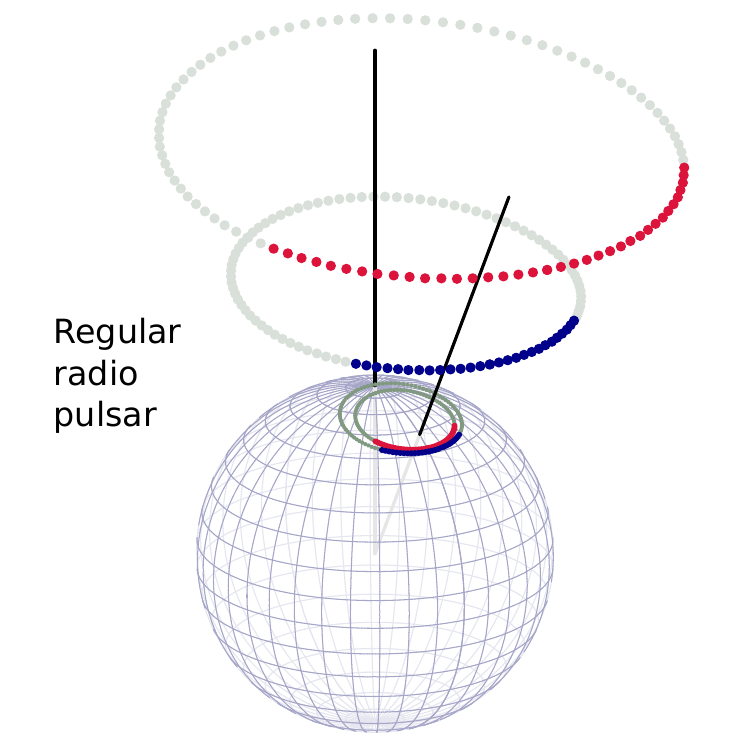}\includegraphics[width=0.25\textwidth]{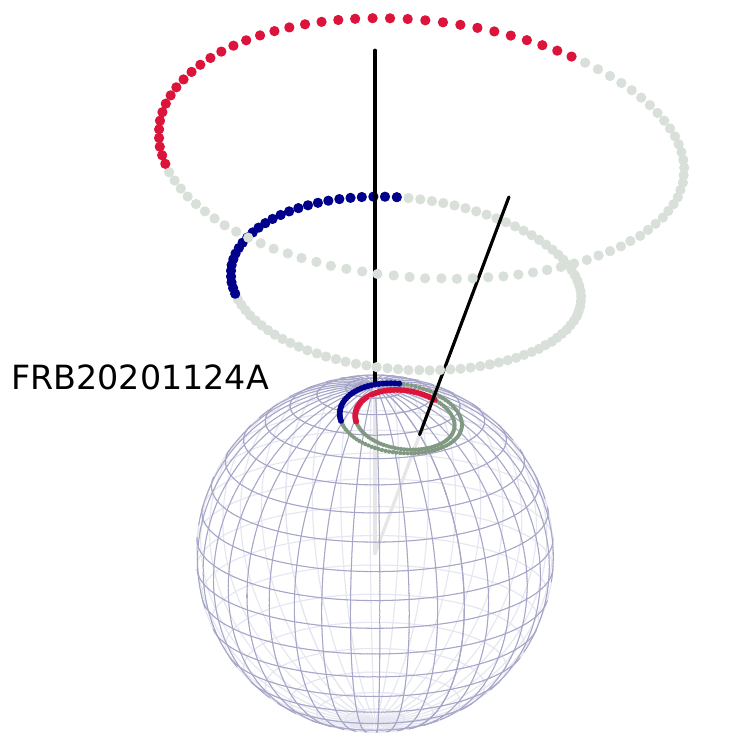}\includegraphics[trim=-10 -10 -10 -10,clip,width=0.25\textwidth]{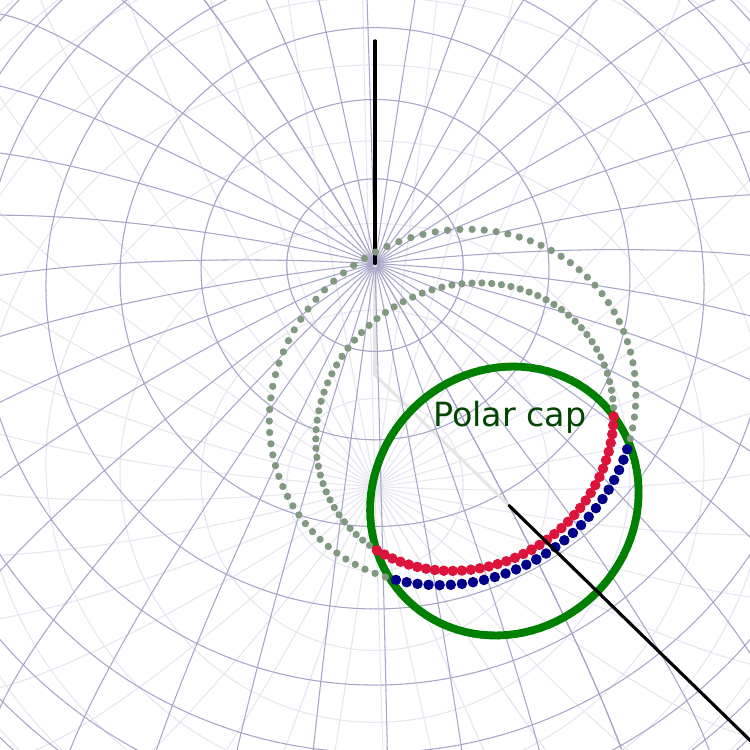}\includegraphics[trim=-10 -10 -10 -10,clip,width=0.25\textwidth]{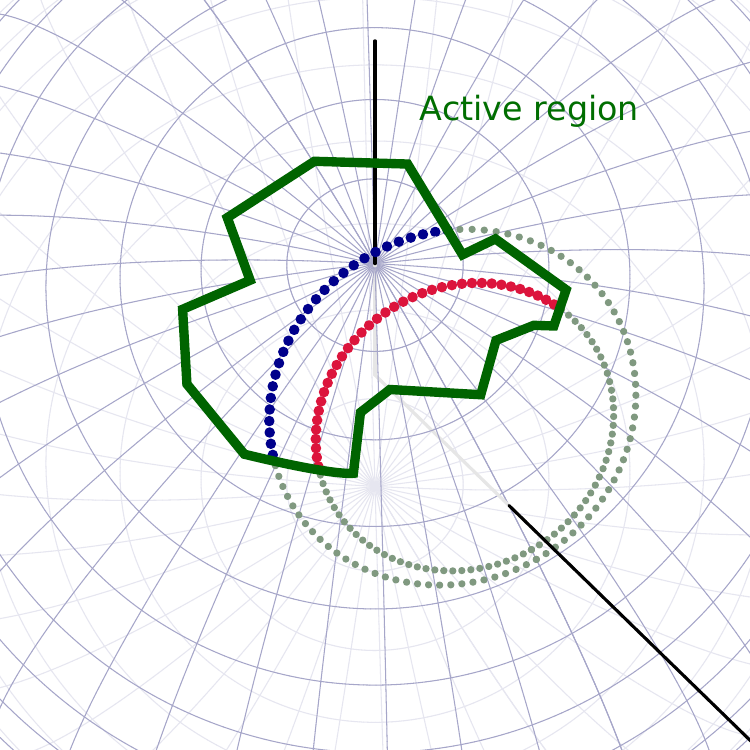}
\caption{Radius-to-frequency mapping of radio emission observed at two radio 
frequencies together with the footpoints of visible emission lines on the 
surface of the star for radio pulsars (left column) and \src\ (right column). 
Spin and magnetic axis are marked with vertical and inclined lines, respectively. 
The path of the LOS and corresponding line footpoints are shown with circle 
markers, gray for no emission, and red (resp. blue) for low (resp. high) frequency. 
For radio pulsars, only open field lines originating within the polar cap 
(green circle) can produce radio emission. At lower frequencies the LOS spends 
more time within the polar cap, resulting in a widening of observed radio profile.
For \src, emission is not restricted to the open field lines and comes from an 
active region of unknown shape. At the end of the \Spring\ {activity window} the LOS crossed 
the edge of an active region (right edge on the figure) in two places, 
with the edge point corresponding to the lower radio frequency extending 
farther along the LOS path. }
\label{fig:RFM}
\end{figure*}

Mathematical expressions for the dipolar magnetic field have the simplest form 
in the rest frame of the pulsar with magnetic moment directed along $z$. In 
this coordinate system the location of the emission region is 
($r_\mathrm{em}$, $\theta_\mathrm{em}$, $\phi_\mathrm{em}$). The relationship 
between ($\theta_\mathrm{ob}$, $\phi_\mathrm{ob}$) and ($\theta_\mathrm{em}$, $\phi_\mathrm{em}$) 
is set by the requirement of the tangent to the magnetic field line at the emission point 
to be aligned with the LOS at the spin phase $\phi_\mathrm{ob}$. Following \citet{Lyutikov2016}:
\begin{equation}
\label{eq:mag2spin}
 \tan \phi_\mathrm{em} = \dfrac{\sin\theta_\mathrm{ob}\sin\phi_\mathrm{ob}}{\cos\alpha\sin\theta_\mathrm{ob}\cos\phi_\mathrm{ob}-\sin\alpha\cos\theta_\mathrm{ob}},
\end{equation}
and
\begin{equation}
 \dfrac{3\cos2\theta_\mathrm{em} + 1}{\sqrt{6\cos2\theta_\mathrm{em}+10}} = \cos\alpha\cos\theta_\mathrm{ob} + \sin\alpha\sin\theta_\mathrm{ob}\cos\phi_\mathrm{ob}.
\end{equation}

The coordinates of the footpoint of an active field line ($\theta_0$, $\phi_0$) 
can be obtained using the equation for that dipolar field line in the rest frame:
\begin{equation}
 \dfrac{R_\mathrm{NS}}{\sin^2\theta_0} = \dfrac{r_\mathrm{em}}{\sin^2\theta_\mathrm{em}},
\end{equation}
and
\begin{equation}
 \phi_0 = \phi_\mathrm{em}.
\end{equation}

Without constraints on the emission altitude $r_\mathrm{em}$, multiple field lines 
can contribute to the emission at any given $\phi_\mathrm{ob}$. 
To  maximize the stellar surface area under consideration in our exploration,
we set the altitude of the observed higher-frequency radio emission to the 
smallest possible value. For the frequency at the center of WSRT band, 
this value corresponding to the minimum altitude from which radio waves
can escape the low-twist magnetosphere. Assuming characteristic values 
of the limiting twist and crust oscillation frequency,
and taking surface magnetic field to be $10^{14}$\,G, we find this minimum escape 
altitude is $5R_\mathrm{NS}$ \citep[\citetalias{Wadiasingh2019},][]{Beniamini2020}.

While our numerical calculations are grounded on the assumption that the observed 
radio waves originate from curvature radiation \citep[][and references therein]{Wang2019}, 
the outcome is qualitatively similar when employing a different relationship 
between the radio wave frequency and the altitude of the emission region. 
For curvature radiation, the relationship between the radio frequency 
and the plasma/magnetic field parameters is 
\begin{equation}
\label{eq:CR}
 \nu = \dfrac{3\pi\gamma^3c}{4\kappa},
\end{equation}
where $\gamma$ is the Lorentz factor of the emitting particles, $c$ is the speed of 
light, and $\kappa$ is the curvature radius of the field lines. For a dipole 
magnetic field, $\kappa$ can be calculated from the coordinates of the
emission point $(r_\mathrm{em}, \theta_\mathrm{em}, \phi_\mathrm{em})$
in the  rest frame of the star:
\begin{equation}
\label{eq:rho_curv}
 \kappa = \dfrac{r_\mathrm{em}(1+3\cos^2\theta_\mathrm{em})^{3/2}}{3\sin\theta_\mathrm{em}(1+\cos^2\theta_\mathrm{em})}.
\end{equation}

Combining Eqs.~\ref{eq:mag2spin}--\ref{eq:rho_curv}, we obtain 
($r_\mathrm{em}$, $\theta_\mathrm{em}$, $\phi_\mathrm{em}$) for both the
WSRT and GMRT radio frequencies $\nu_\mathrm{hi}$ and $\nu_\mathrm{lo}$, 
for a range of chosen inclination and viewing angles. If $\gamma$ does 
not change between two frequencies \citep{Wang2019}, 
$r(\nu_\mathrm{hi})\approx 0.3 r(\nu_\mathrm{lo})$ over the entire range 
of input parameters. This next determines the low-frequency emission to
take place at $16.7R_\mathrm{NS}$, 
above the minimum escape altitude of $15R_\mathrm{NS}$ for GMRT frequencies. 
If $\gamma_\mathrm{lo} < \gamma_\mathrm{hi}$, then $r_\mathrm{lo}$ 
is closer to $r_\mathrm{hi}$, and the lower-frequency $\theta_0(t)$ moves 
closer to its higher frequency counterpart.

The absolute value of $\gamma$ (if taken constant between two frequencies) 
varies from $\sim 50$ to $\sim 200$, with the higher values corresponding 
to field lines closer to the magnetic pole. These values are similar to 
the $\gamma=300$ adopted by \citet{Wang2019}. We note that our values of 
$\gamma$ are lower limits, since we adopted the smallest possible $r_\mathrm{em}$.

%%%%%%%% RVM constraints on geometry
\begin{figure}
\includegraphics[width=0.5\textwidth]{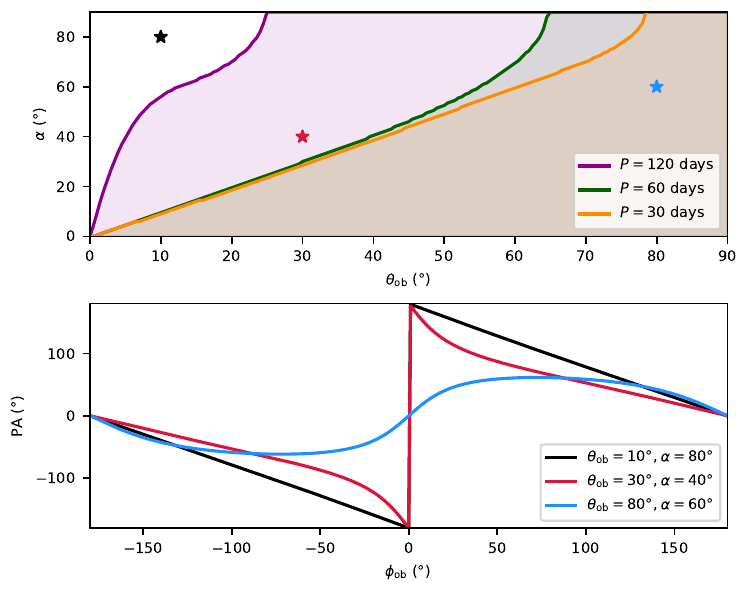}
\caption{\textit{Top:} Constraints on the magnetospheric geometry placed 
by {constant median single-burst PA} during the four-day observing epoch from the  
\Fall\ {activity window} \citep{Jiang2022}. The shaded areas encompass   
inclination--viewing angle pairs related to three trial values of spin 
period $P$ (see legend). For each of these areas, the maximum PA 
{variation} was less than $10\degree$ in at least one of the spin 
phase regions spanning 4\,days/$P$. 
The stars mark three representative geometries whose PA curves are shown in the 
\mbox{\textit{bottom}} plot.
Geometries that approach  aligned rotators and have  small viewing 
angles predominantly show larger PA gradients than allowed
 by the observations, unless the spin period exceeds several months.}
\label{fig:RVM}
\end{figure}

In the absence of any other constraints, the footpoints of the field 
lines with emission visible to the observer form a closed loop 
on the star surface (Fig.~\ref{fig:LOS_on_surface}). 
For the chosen frequencies $\nu_\mathrm{hi}$ and $\nu_\mathrm{lo}$, 
loops subtend $\lesssim 0.2\%$ of star surface for small 
$\theta_\mathrm{ob}$, regardless of inclination angle. The fraction 
of the stellar surface encircled by this footpoint loop increases with 
growing $\theta_\mathrm{ob}$, reaching about 20\% for large viewing 
and small inclination angles. 

Since the footpoint loops are closed, there exists a field line that 
can potentially provide radio emission at any spin phase $\phi_\mathrm{ob}$.
Whether this line will actually emit is determined by physical requirements for 
pair production and radio wave generation. For radio pulsars, such a 
requirement is that radio emission is produced by the open field lines 
with $\theta_0\leq\theta_\mathrm{PC}$, where $\theta_\mathrm{PC}$ is the 
radius of the polar cap. Figure~\ref{fig:RFM} provides an illustration of this, 
showing an example of the footpoint loops at two radio frequencies for an 
inclination angle of $\alpha=30\degree$ and $\theta_\mathrm{ob} = 60\degree$. 
For plotting convenience, $r_\mathrm{lo}$ was set to be $2R_\mathrm{NS}$ and 
$\nu_\mathrm{hi} = 1.5\nu_\mathrm{lo}$. The polar cap with 
$\theta_\mathrm{PC} = 15\degree$ cuts segments from the footpoint loops, with 
the lower-frequency segment spanning a larger amount of spin phase bins. 
This implies that radio emission at lower frequencies is observed at earlier 
and later $\phi_\mathrm{ob}$ than the high-frequency emission;  the average 
pulse profile is wider at lower frequencies.

In the \citetalias{Wadiasingh2019}  model the  \src\  emission 
can originate on closed field lines, eliminating the polar cap requirement. 
%\footnote{Polar cap radius 
%would be on the scale of 10\,cm for a month-long spin period}. 
The minimum emission altitude requirements imposed by the plasma transparency 
set limits on the magnetic colatitude $\theta_0$, defined by 
$r_\mathrm{min, alt}=R_\mathrm{NS}\sin^2\theta_0$, indicating that 
$\theta_{0}\lesssim15\degree$ and $26\degree$ for lower and higher 
frequencies, respectively. This broad region may host several active 
patches during the {activity window}.

A frequency-dependent activity edge arises when the LOS crosses the edge 
of the active patch. The exact lag between high- and low-frequency 
quenching is determined by the skewness of the active region shape 
in the magnetic longitude direction and the unknown value of the 
spin period. For instance, assuming the \src\  spin period to be 2 months 
(the duration of the \Spring\ {activity window}), then 
$\delta \phi_\mathrm{ob} = 20\degree$--$40\degree$ for a 3--6 day lag, 
which translates to a similar upper limit on $\delta \phi_\mathrm{em}$, 
with the latter being smaller for non-concentric footpoints loops. 
Thus, a small variation in active region shape may lead to a large 
variation in observed cessation lag.

A frequency-dependent activity window places constraints on the overall 
size of an active surface area. Individual crustal motions, on the other hand,
are limited in size by the trombone frequency drift
\citep{Wang2019,Bilous2022,Lyutikov2020}.
Such frequency drift by $\delta\nu = 300$\,MHz at 1400\,MHz or by 125\,MHz 
at 650\,MHz would translate to 
$\delta \theta_0 \approx 0.5\theta_0\delta \nu/\nu\approx 0.1\theta_0\lesssim3\degree$.
If the size of the active ``spark'' which produces an individual 
FRB is comparable to the amplitude of footpoint dislocation $\xi$, 
then the latter is less than 500\,m, below $\xi_\mathrm{max}=3$\,km 
from \citet{Beniamini2020}.

\subsection{The rotating vector model (RVM)}
\label{sec:geo:rvm}

In the rotating vector model, the plane of the linearly polarized part 
of the radio emission is defined by the field line curvature  plane at 
the emission point. For an inclined dipolar field, the position angle (PA)
has a characteristic S-shape dependence on the spin phase $\phi_\mathrm{ob}$
and the angles introduced  previously, as defined by the following analytical relation:
\begin{equation}
\label{eq:RVM}
 \tan\mathrm{PA} = \dfrac{\sin\alpha\sin\phi_\mathrm{ob}}{\sin\theta_\mathrm{ob}\cos\alpha-\cos\theta_\mathrm{ob}\sin\alpha\cos\phi_\mathrm{ob}}.
\end{equation}

For both normal radio pulsars and magnetars, individual pulses exhibit 
a large diversity of PA behavior \citep{Mitra2016,Johnston2024,Kramer2007}. 
While numerous deviations are known \citep{Johnston2023}, the measured 
single-pulse PAs average to an S-curve defined
by Eq.~\ref{eq:RVM} in a considerable number of cases. 

If we refocus now on {\src}, we see that 
the behavior of the PA curve is known only for a four-day stretch at
the end of the \Fall\ {activity window} \citep{Jiang2022}. 
{There, the collection of the mean PAs of individual bursts had a 
relatively wide distribution around some most probable value, which in turn 
remained constant during four observing days.}
The authors do not specify the limits of the 
constancy, but the distribution of mean PAs in  \citet{Jiang2022} has a standard deviation 
of about $20\degree$. For the purpose of the  arguments  following below,
we adopt a PA-gradient upper limit of $10\degree$ per four days 
as best describing the observations.

We proceed by selecting ($\alpha$, $\theta_\mathrm{ob}$) pairs for which 
there exists at least one spin longitude window that can produce the observed 
$|\max(\mathrm{PA})-\min(\mathrm{PA})\,| < 10\degree$. The size of the 
spin window depends on the unknown spin period, so we explore three 
trial periods, of one, two, or four months. Figure~\ref{fig:RVM} shows 
the allowed ranges of inclination/viewing angles. For the smallest trial 
period of one month $\alpha \lesssim \theta_\mathrm{ob}$ unless the inclination
angle is close to $90\degree$. This excludes the non-concentric types of footpoint 
loops that are visible in Fig.~\ref{fig:LOS_on_surface}. The inclination angle becomes less
constrained as the potential spin period grows. Combinations of large $\alpha$
and small $\theta_\mathrm{ob}$ (e.g., $\alpha=80\degree$ and 
$\theta_\mathrm{ob}=10\degree$) still, however, produce a PA gradient that is too steep, even 
if $P = 4$\,months.

\section{Summary and conclusions} \label{sect:summary}

\src\ is a prolific source of FRBs. Its bursts exhibit rich and complex structure 
and are undoubtedly capable of providing valuable insights into the underlying 
emission physics and propagation mechanisms. However, the current body of 
observational evidence contains gaps in areas of high scientific interest, 
and is inevitably affected  by both instrumental and processing limitations. 

\subsection{Determining DM and fluence changes requires careful analysis}

One example of these limitations is our current inability to disentangle 
the dispersive delay imposed by the interstellar medium from the intrinsic 
burst properties. We have demonstrated that the DM measured using
the de facto standard structure-maximizing method absorbs the trombone 
drift for fainter pulses, leading to an overestimation bias in DM measurements that can 
be as large as 10\,\dmu,  even for bursts with an integrated 
S/N on the order of 100.  For brighter bursts, the complex structure often
 cannot be maximized with a single DM value \citep[e.g.,][]{Zhou2022}.
This bias should be taken into account while analyzing DM distributions 
aggregated from several studies with different sensitivities since 
it can lead to apparently  multi-modal DM distributions, which can 
subsequently be interpreted as plasma lensing or frequency-dependent 
DM \citep{Wang2023b}. 
On the other hand, directly incorporating external constraints on 
spectro-temporal FRB behavior would
allow for testing interesting physical effects. One example is investigating
the filamentation of FRBs in the relativistic winds of magnetars, which implies
a dependence of the measured DM on pulse luminosity due to propagation
in the near-source environment \citep{Sobacchi2023}. 

We confirm the absence of secular DM trends between different activity 
windows of \src\, down to the level of a few \dmu. This places a limit on 
any electron density changes along the LOS that could  have been caused by an active 
circum-burst environment \citep{Metzger2017}. 
Combining WSRT bursts from the \Winter\ {activity window} and burst 
G05 from the \Spring\ {activity window},
we obtain a broadband PL dependence of the decorrelation bandwidth,   
shallower  than expected for Kolmogorov turbulence and close to the 
measurements of \citet{Main2022}, who had a similar observing setup.

Despite the relative abundance of available data, the \src\ burst-fluence 
distribution remains poorly constrained. The FAST 
observations demonstrate that there is a significant amount of variability in the 
fluence distribution over a range of timescales: 
 between subsequent days to  between distinct 
activity windows \citep{Xu2021,Zhang2022}. Corroborating  the findings of 
\citet{Kirsten2024}, we observed a shallower PLI for bursts 
in the \Winter\ {activity window} than in the FAST measurements 
during previous windows (all at   frequencies around   1300\,MHz).  However, as we demonstrate, 
PLI measurements tend to be imprecise and biased for small burst samples, 
especially when drawn from distributions with flattening at the low-fluence end, 
even if the observed distribution appears to be  well fit  with a single PL.

\subsection{A ULP magnetar model for the burst behavior}

We explored the possibility that  \src\ is an ultra-long-period 
magnetar with a spin period on the scale of months, a period that  may remain undetectable 
for current searches. For such ULP sources, FRBs could be triggered by motion in 
the star crust, if these cause dislocation of magnetic field-line footpoints 
that leads to the production of plasma, which subsequently emits radio waves
via some pulsar-like emission mechanism (\citetalias{Wadiasingh2019}). 
Similar to the short X-ray bursts thought to be triggered by similar crustal
motion events, the arrival times of individual FRBs would then follow a 
log-uniform distribution if measured from the trigger moment.
Such a distribution has been reported for the repeating source \srcRi\ (\citetalias{Wadiasingh2019}). 
We demonstrated that TOAs of individual bursts from  \src\ 
in both the \Spring\ and \Fall\ FAST samples exhibit a distribution that 
is uniform, not log-uniform. The sample could, however,
still actually be part of the tail of a log-normal distribution, from 
a trigger event that occurred before these FAST observing sessions. 
As the burst rate, in a crescendo,  increases  exponentially before the emission 
quenches at the end of the \Fall\ {activity window} \citep{Zhang2022}, 
there must be multiple trigger events during a single {activity window}.
If these occurred several hours before the observation,
our data is consistent with this model.

Another prediction from the crust motion and low-twist theory links crustal oscillations to 
quasi-periodic sub-burst spacing within individual FRBs. Our WSRT sample contains 
three bursts with seemingly periodic components; however, we were unable to 
find any significant periodicity. We show (and caution) that detecting periodicity 
for closely spaced sub-bursts of arbitrary shape is difficult, 
especially when the number of period trials is unconstrained.

If radio emission originates from only one frequency-dependent range of heights, 
then at any given spin phase, the LOS samples only a small fraction of the stellar 
surface. In contrast, X-ray bursts can be detected from the entire stellar surface 
at any moment. This difference may explain the intriguing result obtained by 
\citet{Tsuzuki2024}, who found that regular radio pulses from the occasionally 
FRB-emitting magnetar SGR 1935+2154 exhibit correlation properties similar to 
those of extragalactic FRBs \citep{Totani2023} and earthquakes. However, 
X-ray bursts from the same source did not show such correlation properties.
We hypothesize that the lack of correlation for X-ray bursts stems from the 
superposition of several starquake events occurring simultaneously at different 
locations. In radio, individual starquake events are observed sequentially.

\subsection{The {\src} {activity window} ends  chromatically}

At the end of the \Spring\ {activity window}, our GMRT observations recorded several strong 
bursts at 300--600\,MHz. This is surprising because only  a few hours before,
FAST had placed upper limits on 
the absence of emission at 1250\,MHz that were 1000 times more stringent. 
By combining burst detections and telescope scheduling information, we 
were able to paint a complete multi-frequency picture of the end of this {activity window}.
We  showed  that the edge of the \Spring\ {activity window} depends on the 
observing frequency:  low-frequency emission is present for 3--6\,days 
after the higher-frequency quenching. This marks the second detection of 
a chromatic {activity window} for a repeating FRB source, following 
\srcRiii\ \citep{Pastor-Marazuela2021,Pleunis2021b}. However, for 
\srcRiii, a 16.3-day periodicity is known, and the chromatic window 
was established by combining bursts from several periods, whereas 
for \src, we have so far recorded only a one-time event. Unfortunately,
the end of the \Fall\ and \Winter\ {activity window}s was not 
covered by multi-frequency observations.

The fact that lower-frequency emission was recorded a few days 
after the disappearance of the higher-frequency emission would be hard to explain within 
the plasma lensing theory of burst quenching. In this theory, the non-uniform 
distribution of free electrons along the LOS acts as a diverging lens, causing 
a significant increase in burst rate followed by a sudden drop
at the end of the \Fall\ {activity window} \citep{Chen2024}. For the 
one-dimensional Gaussian lens model reviewed by \citeauthor{Chen2024},
low-frequency emission disappears before the high-frequency emission.
This is the opposite of what we observed.

\subsection{Two well-known pulsar models can explain the chromatic activity end, and the flat polarization,  assuming a rotation period of order one month}

The chromatic {activity window} we observed  resembles a long-known effect 
observed for radio pulsar emission, namely the widening of the on-pulse 
window toward lower frequencies. This similarity can actually have a physical 
interpretation if  FRBs are indeed generated via a pulsar-like emission 
mechanism. We applied the classical phenomenological
radius-to-frequency model of average 
profile widening to the frequency-dependent edge of the \Spring\ 
{activity window}. Our goal was to constrain the magnetospheric geometry, 
and the location of active regions on the star surface. Assuming 
that the active regions are not confined to the open field lines, and 
given the fact that the spin period is unconstrained, we find a 
multitude of possible active regions for any combination of dipole 
inclination angle and the viewing angle of the LOS. The possible extent of these 
regions in magnetic colatitude is only limited by the requirement of 
plasma transparency and is on the order of $\sim 20\degree$. 
The observed chromatic edge of the {activity window} is defined by the 
exact shape of the active regions, and small variations in
this shape may cause different amounts of chromatic lag. If active regions 
differ from one {activity window} to another, we expect the amount of 
frequency-dependent quenching timescale to vary in subsequent {activity window}s as well.
In this model an inverted dependency (i.e., low-frequency emission disappearing before 
the high frequencies) is also allowed. Similarly, a chromaticity
at the onset of the activity might also exist, again depending on the
exact shape of the active regions. In the absence of a physical 
constraint on the active region shapes, all possibilities, such as
symmetric, asymmetric, or one-sided  behavior  of the
{activity window} would be possible. 
It is worth noting that \srcRiii\ exhibits such asymmetric chromaticity,
with the emission window at 5\,GHz preceding the one at 150\,MHz \citep{Bethapudi2023}.

The magnetospheric geometry of \src\  can be  further constrained
using a second classical aspect of our understanding of pulsars,
the rotating vector model. For this, we determined which geometries are 
allowed under the observed PA constancy during the four-day observing 
stretch at the end of the \Fall\ {activity window} \citep{Jiang2022}. 
For potential spin periods of a month and above, we can exclude aligned 
rotators with a LOS close to the spin axis. The constraints are less 
strict  for larger potential spin periods.

\subsection{Forward look}

To the best of our knowledge, as of July 2024 there have been no 
indications that  \src\ entered a new {activity window}, with the
last pulses detected in the end of March 2022 \citep{Wu2024}.
Continuous monitoring of this source remains critically important
by, among others, CHIME on the low-frequency end and a collection of 
small telescopes operating at higher frequencies \citep{Ould-Boukattine2024}.

We showed that catching a new {activity window} early is essential for 
testing the crust motion and low-twist FRB emission theory, as future observations may 
unveil a prolific cluster of bursts that could mark the onset of a 
single trigger event. Additionally, such observations could provide 
valuable insights on the potentially diverse shape of the active regions 
through the recording of another chromatic edge of the {activity window},
after the first one presented here. 
Finally, further constraints on the secular behavior of the PA 
would be helpful in refining our understanding 
of magnetospheric geometry.

\begin{acknowledgements}
AB thanks Matteo Trudu, Adam Lanman, Jian-Ping Yuan, and Franz Kirsten for 
providing extra information about \src\ observations, Kejia Lee 
for help with downloading FAST data, Zorawar Wadiasingh and 
Andrey Timokhin for discussions on the low-twist ULP magnetar model, and Heng Xu for
providing critical feedback on fluence distributions.
This work was supported by the 
European Research Council under the European Union's Seventh Framework 
Programme (FP/2007-2013)/ERC Grant Agreement No. 617199 (`ALERT') and Vici 
research programme `ARGO' with project number 639.043.815, financed by 
the Dutch Research Council (NWO). 
JVL further acknowledges support from 
'CORTEX' (NWA.1160.18.316), under the research programme NWA-ORC by NWO.
YM further acknowledges support from the Department
of Science and Technology, India, via the Science and Engineering Research Board
(SERB) Start-up Research Grant (SRG/2023/002657). IPM further acknowledges 
funding from an NWO Rubicon Fellowship, project number 019.221EN.019. 
We would like to thank the Centre Director, NCRA-TIFR, and the 
GMRT observatory for the prompt director's discretionary time 
allocation and scheduling of our observations. We thank the staff of 
the GMRT who have made the GMRT observations possible.
The GMRT is run by the National Centre for Radio Astrophysics of the
Tata Institute of Fundamental Research.
This work makes use of data from the Apertif system installed at the 
Westerbork Synthesis Radio Telescope owned by ASTRON. ASTRON, 
the Netherlands Institute for Radio Astronomy, is an institute of NWO.\vspace{1ex}\\

\textit{Code availibility.} 
An ipython notebook that implements the MLE (and other) power-law fits and 
includes further instructions is hosted at 
\url{https://doi.org/10.5281/zenodo.12644702}
and \url{https://github.com/TRASAL/FRB_powerlaw}.
\end{acknowledgements}

\bibliographystyle{aa} 
\bibliography{bibliography} 

%\clearpage
\appendix

\section{Comparing least-squares and MLE methods of PLI estimates}
\label{app:PL_fit}

%%%%%%%% Large underestimate of errors by least square method
\begin{figure}
\includegraphics[width=0.45\textwidth]{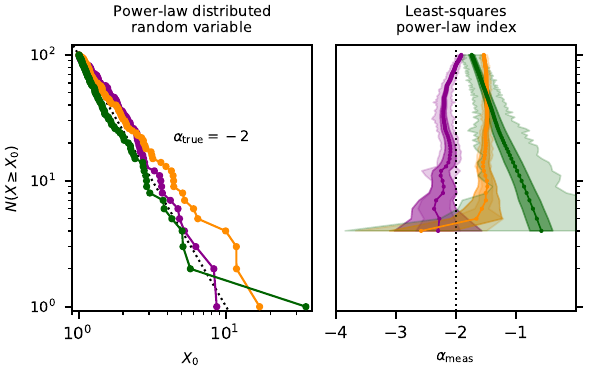}
\caption{\textit{Left:} Survival functions for three $N=100$ samples of 
simulated random variables distributed according to Eq.~\ref{eq:sf_pl} 
with $\at = -2$. \textit{Right:} PLI $\am$ obtained with the least-squares 
method for the subsamples consisting of $4\leq N\leq 100$ elements sorted in 
decreasing order. The shaded regions mark the standard least-squares 
error (darker shade) and the bootstrapping error (lighter shade).}
\label{fig:PL_graph}
\end{figure}

%%%%%%% Accuracy of three PL fitting methods
\begin{figure}
\includegraphics[width=0.5\textwidth]{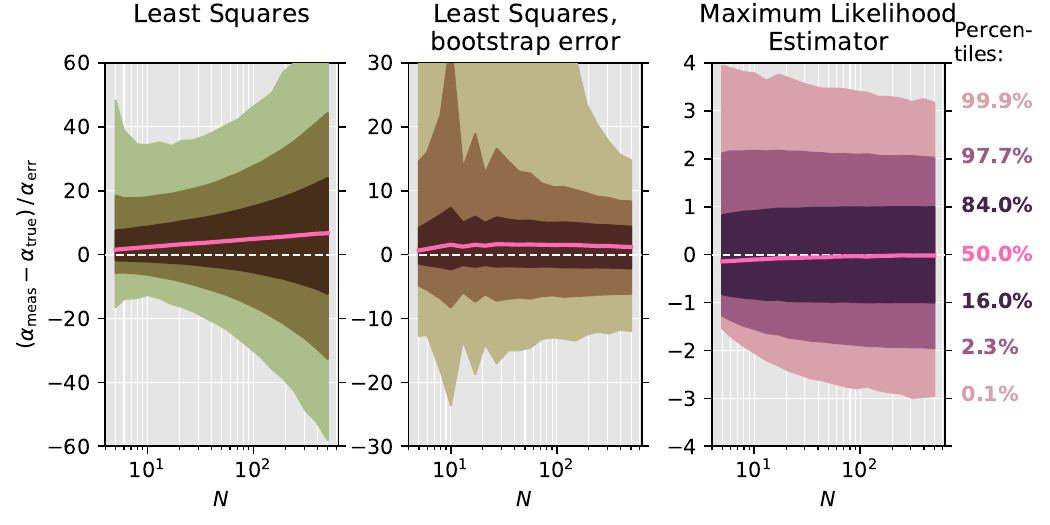}
\caption{Accuracy of three methods of $\alpha$ estimation vs sample size $N$.
\textit{Left:} Graphical method with its least-squares error;
\textit{center:} Graphical method with a bootstrap error;
and \textit{right:} Unbiased MLE method according to 
Eqs.~\ref{eq:a_unbiased}--\ref{eq:a_err_unbiased}.
We note that the y-axis scale decreases by over a factor of 10 from left to right. 
The pink line corresponds to the median of 
$(\am-\at)/\alpha_\mathrm{err}$, and the shaded regions mark percentiles 
corresponding to $\pm 1$, $\pm2$, $\pm3$ standard deviations of 
the normal distribution.}
\label{fig:PL_accuracy}
\end{figure}

%%%%%%% Accuracy of fitting with powerlaw package
\begin{figure}
\includegraphics[width=0.5\textwidth]{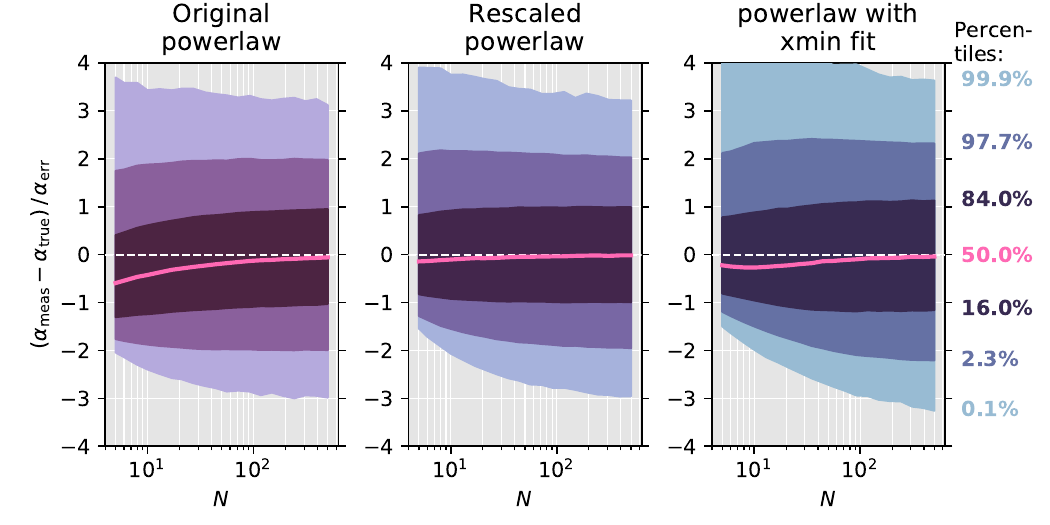}
\caption{Similar to Fig.~\ref{fig:PL_accuracy}, but for accuracy of $\alpha$
estimates using the \texttt{powerlaw} package with different modifications.
\textit{Left:} Original package version with biased estimates 
(Eqs.~\ref{eq:a_biased}--\ref{eq:a_err_biased}) and fixed $X_\mathrm{min}$; 
\textit{center:} Unbiased estimates with fixed $X_\mathrm{min}$; and \textit{right:}
Unbiased estimates with fitted $X_\mathrm{min}$. In all three cases a random variable was simulated with single-PL distribution according to Eq.~\ref{eq:sf_pl}.}
\label{fig:PL_accuracy_powerlaw}
\end{figure}

In this section we   investigate accuracy and bias of PLI estimates provided by 
the graphical and MLE methods. We start with demonstrating the sub-optimal 
performance of the least-squares estimates. To do this, we generated three 
realizations of PL-distributed random variables using a Pareto distribution 
from the python library \texttt{scipy}, with survival function defined as:
\begin{equation}
 N(X\geq X_0) = \Ns\left(\dfrac{X_0}{X_\mathrm{min}}\right)^\alpha.
 \label{eq:sf_pl}
\end{equation}
We took $\Ns=100$, $X_\mathrm{min}=1$, and $\alpha=-2$. PLIs were estimated 
using graphical method on the unbinned survival functions. In order to 
investigate the dependence of the PLI produced by the graphical method
on the sample size, we performed  fits on reduced samples, selecting 
$X\geq X_\mathrm{min}$, where 
$X_\mathrm{min}$ ranged from the smallest to the fourth largest values 
of $X$. For each fit both the least-squares and bootstrapping errors 
were estimated. For the latter we followed the procedure from \citet{Kirsten2024}, 
removing 10\% of the sample without replacement (or one element 
if $\Ns(X_\mathrm{min})<10$), calculating the PLI, repeating this 100 times and 
then taking standard deviation of the acquired distribution as $\epsilon_\alpha$,
the  error on $\alpha$. Figure~\ref{fig:PL_graph} displays
the survival functions and the resulting $\alpha(\Ns)$. Even for $\Ns=100$,
the resulting PLI can deviate from the true value by as much as 25\%, 
while the formal errors are an order of magnitude smaller.

The uncertainties in PLI measurements calculated using the bootstrapping 
method are indicative at best. The distribution of $\am$ on bootstrapped 
subsamples is generally skewed toward shallower PLIs. 
The shape of the SF tail can be heavily influenced by a few large random variables, 
which constitute a small fraction of the entire sample. Depending on the chance 
presence of these variables in the bootstrapped subsample, the measured PLI can
vary substantially, resulting in multi-modal bootstrapped PLI distributions. 
This has a significant effect on the
calculated standard deviation of bootstrapped PLIs, causing large scatter in the
bootstrapped error on $\am$ from one simulation to another. At the same time, 
the error is not large enough to describe the true discrepancy between $\at$ 
and $\am$ (Fig.~\ref{fig:PL_graph}).

To further test the quality of the PL fitting, we examined the distributions 
of test statistics $a$, $a = (\am-\at)/\alpha_\mathrm{err}$ for several fitting methods. We generated a series of samples with $N$ ranging from 5 to 500, 
values for  $\at$ and $-2$. Performing smaller tests using different 
values of $\at$ did not reveal any differences in the behavior of examined distributions. 

Fig.~\ref{fig:PL_accuracy} shows the accuracy of the fit for three methods: 
graphical with its standard errors, graphical with the bootstrap errors, and MLE 
(where we emphasize here that the three relevant ordinates in that figure are on different scales).
Following \citet{James2019}, we used unbiased $\alpha$ estimates:
\begin{equation}
 \dfrac{1}{\am} = \dfrac{1}{\Ns-1} \sum\mathrm{ln}\left(\frac{X}{X_\mathrm{min}}\right).
 \label{eq:a_unbiased}
\end{equation}
For $\alpha_\mathrm{err}$ we used the expression from the same work, despite 
it being valid only for large $\Ns$:
\begin{equation}
 \alpha_\mathrm{err} = \dfrac{\am}{\sqrt{\Ns-2}}.
 \label{eq:a_err_unbiased}
\end{equation}

In the ideal case $a$ follows a normal distribution with a median of 0 and a standard
deviation $\sigma=1$. This means that $\am$ provides an unbiased estimate of $\at$, 
and $\alpha_\mathrm{err}$ has an intuitive Gaussian meaning. To illustrate 
how closely $a$ obeys a normal distribution we 
plot, in Fig.~\ref{fig:PL_accuracy} (right sub-panel) , 
the percentiles at levels corresponding to the median and its $\pm \sigma$, 
$\pm 2\sigma$, and $\pm 3\sigma$ counterparts.
For the MLE method and relatively large sample sizes $N\gtrsim 100$,
these levels are at $a = 0, \pm 1, \pm 2, \pm 3$, indicating that 
 $a$ very closely resembles the normal distribution. For smaller 
sample sizes $a$ is skewed toward negative values, and accurately 
described by a gamma distribution (Eq.~\ref{eq:post_prob}). We note that the median value of $a$ stays close to 0 for all sample sizes.

In stark contrast to this, the graphical method leads to a median value of $a$ 
on the order of 5 for the least-squares error and 1.5 for the bootstrap error, 
indicating that median $\am$ tends to be larger than $\at$ by a few 
$\alpha_\mathrm{err}$, regardless of sample size. The wide spread of the percentile curves 
in Fig.~\ref{fig:PL_accuracy} (left and middle sub-panel) 
demonstrates that $\alpha_\mathrm{err}$ values are substantially underestimated 
when choosing to use the graphical method .

We performed similar accuracy estimates for PLIs determined with the Python package \texttt{powerlaw}\footnote{\url{https://pypi.org/project/powerlaw/}} 
\citep{Alstott2014}. In the original code, $\alpha$ and $\alpha_\mathrm{err}$
are calculated as 
\begin{equation}
 \dfrac{1}{\am} = \dfrac{1}{\Ns} \sum\mathrm{ln}\left(\frac{X}{X_\mathrm{min}}\right),
 \label{eq:a_biased}
\end{equation}
and
\begin{equation}
 \alpha_\mathrm{err} = \dfrac{\am}{\sqrt{\Ns}}.
 \label{eq:a_err_biased}
\end{equation}
Fig.~\ref{fig:PL_accuracy_powerlaw} (left panel) shows that this estimate performs worse 
than the version modified according to Eqs.~\ref{eq:a_unbiased}-\ref{eq:a_err_unbiased} 
(center panel).

In real-life observations $X_\mathrm{min}$ is often not known. The \texttt{powerlaw} 
package offers fitting for the best $X_\mathrm{min}$ by minimizing the 
Kolmogorov-Smirnov distance between the data and the theoretical 
power-law fit. We have repeated the aforementioned simulations with 
fitting for $X_\mathrm{min}$, restricting its range from $\mathrm{min}(X)$ 
to the third largest value. In this case, the median value of 
$a$ is slightly more biased toward small negative 
values and the distribution is somewhat wider (Fig.~\ref{fig:PL_accuracy_powerlaw}), 
but the MLE still provides a much better fit than the graphical method.

%\software{RFIClean \citep{Maan2021}, PRESTO \citep{RansomThesis}, SIGPROC, DSPSR \citep{vSB11}}
%%\facility{GMRT(GWB)}

\end{document}